%% file: qclock_arXiv.tex
\documentclass[aps,pre,amsmath,amssymb,10pt,a4paper,floatfix,twocolumn,superscriptaddress]{revtex4-1}

\usepackage{titlesec}
\usepackage{bm}
\usepackage{comment}
\usepackage{verbatim}

\usepackage{graphicx}
\usepackage{subfigure}
\usepackage{tabularx}

\input{preamble.tex}

\usepackage[colorlinks,bookmarks=true,citecolor=blue,linkcolor=red,urlcolor=blue,citecolor=blue]{hyperref}
\def\be{\begin{equation}}
\def\ee{\end{equation}}
\def\bea{\begin{eqnarray}}
\def\eea{\end{eqnarray}}

\def\vec{\mathbf}

\usepackage{times}

\begin{document}

\title{Critical properties of the two-dimensional $q$-state clock model}

\date{\today}

\author{Zi-Qian Li}
\affiliation{Institute of Physics, Chinese Academy of Sciences, Beijing 100190, China}
\affiliation{University of Chinese Academy of Sciences, Beijing 100049, China}
\author{Li-Ping Yang}
\affiliation{Department of Physics, Chongqing University, Chongqing 401331, China}
\author{Z. Y. Xie}
\affiliation{Department of Physics, Renmin University of China, Beijing 100872, China}
\author{Hong-Hao Tu}
\email{hong-hao.tu@tu-dresden.de}
\affiliation{Institute of Theoretical Physics, Technische Universit{\"a}t Dresden, 01062 Dresden, Germany}
\author{Hai-Jun Liao}
\email{navyphysics@iphy.ac.cn}
\affiliation{Institute of Physics, Chinese Academy of Sciences, Beijing 100190, China}
\affiliation{Songshan Lake Materials Laboratory, Dongguan, Guangdong 523808, China}
\author{T. Xiang}
\email{txiang@iphy.ac.cn}
\affiliation{Institute of Physics, Chinese Academy of Sciences, Beijing 100190, China}
\affiliation{University of Chinese Academy of Sciences, Beijing 100049, China}
\affiliation{Collaborative Innovation Center of Quantum Matter, Beijing 100190, China}

\begin{abstract}
We perform the state-of-the-art tensor network simulations directly in the thermodynamic limit to clarify the critical properties of the $q$-state clock model on the square lattice. We determine accurately the two phase transition temperatures through the singularity of the classical analog of the entanglement entropy, and provide extensive numerical evidences to show that both transitions are of the Berezinskii-Kosterlitz-Thouless (BKT) type for $q\ge 5$ and that the low-energy physics of this model is well described by the $\mathbb{Z}_q$-deformed sine-Gordon theory. We also determine the characteristic conformal parameters, especially the compactification radius, that govern the critical properties of the intermediate BKT phase.
\end{abstract}

\maketitle
{\em Introduction}. 
The idea of utilizing simple toy models to understand complex phenomena lies at the heart of physics. The practice of this guiding principle has achieved particular success in the study of continuous phase transitions. In the critical regime, the correlation length is infinite, and the system becomes scale invariant. Critical phenomena are described by field theory in the long wavelength limit, and their physical properties are governed by universal critical exponents. Two celebrated examples include the Landau-Ginzburg-type continuous phase transitions~\cite{LandauBook} driven by fluctuating local order parameters with symmetry breaking and the Berezinskii-Kosterlitz-Thouless (BKT) transitions~\cite{Berezinskii,KT1,KT2,KT3} driven by topological defects (vortices).

A simple toy model that exhibits many of these fascinating features is the so-called $q$-state clock model, which is a discretized spin $XY$ model, defined on the square lattice. The Hamiltonian reads
\begin{equation}
H = -\sum_{\left\langle ij \right\rangle}\cos(\theta_i-\theta_j),
\label{Eq:qclock_Model}
\end{equation}
where the $q$ spin states at site $i$ are denoted by a planar angle $\theta_{i}=2 \pi k_i / q$ with $k_i = 1, \ldots, q$, and $\langle ij \rangle$ stands for the nearest neighbors. The coupling is ferromagnetic. In the case $q=2,3,4$, this model is exactly soluble. It reduces to the Ising and $\mathbb{Z}_3$ Potts models when $q=2$ and $3$, respectively. The $q=4$ model is equivalent to two copies of the Ising model. For these three cases, the system exhibits a second-order phase transition from a high-temperature paramagnetic phase to a low-temperature ferromagnetic ordered phase. In the limit $q \rightarrow \infty$, the discrete spin approaches
a planar rotor, and Eq. (\ref{Eq:qclock_Model}) reduces to the standard $XY$ model.
In this limit, there is no spontaneous symmetry breaking, but there is still a phase transition between the low-temperature vortex-antivortex condensed BKT phase and the high-temperature paramagnetic phase. Thus the $q$-state clock model offers a unique and unified playground to understand both the second-order Landau-Ginzburg and infinite-order BKT transitions.

In past decades, the $q$-state clock model has been extensively investigated both analytically~\cite{Kramers1941,FYWuq5,Alcaraz1980,
FYWu,Wiegmann1978,Elitzur,Jose,Cardy,Nomura1995,Nomura2002,Matsuo2006,BA_duality} and numericallly
~\cite{Tobochnik1982,challa,yamagata,tomita,Lapilli2006,Baek2009,brito,Borisenko2011,Borisenko2012,baek2013,
kumano,Negrete2018,Chatterjee2018,Surungan2019,DMRG_q5,nishino_q6,JingChen,yujifeng2018,hwang,Kim2017,
Hong2019}.
Nevertheless, the nature of the two transitions to the intermediate critical phase from either the high-temperature paramagnetic or the low-temperature ferromagnetic phase, more specifically whether they are of BKT-type for small $q$, such as $q=5$ and $6$, are elusive. Accurate determination of the two critical temperatures for the cases $q>4$ also remains a challenging problem (see Table~\ref{Table:Tc_List} in Supplemental Material (SM)). This model is known to be self-dual under the Kramers-Wannier dual transformation for $q=2,3,4,5$~\cite{Kramers1941,Alcaraz1980,JingChen}. However, it is not clear whether there is a self-dual point for an arbitrary $q$.

In this work, we employ the state-of-the-art tensor network methods to address these open issues. The tensor network calculations, performed directly in the thermodynamic limit, allow us to analyze with high precision the critical behavior of the model for the cases $q\ge 5$. In particular, we find that a singularity emerges in the classical analog of the entanglement entropy of the fixed-point matrix-product-state (MPS)~\cite{nishino_q6}, which offers an accurate approach for us to determine the two critical temperatures. We have examined the effective low-energy field theory of the $q$-state clock model by comparing the numerical calculations with the theoretical predictions, and evaluated the conformal coupling parameter, compactification radius, of the $\mathbb{Z}_q$-deformed sine-Gordon model, using the Klein bottle entropy
approach~\cite{Tu2017,Tang2017a,Chen2017a,Wang2018,Tang2019a}.

\begin{figure}[t]
\centering
\includegraphics[width=7.0cm]{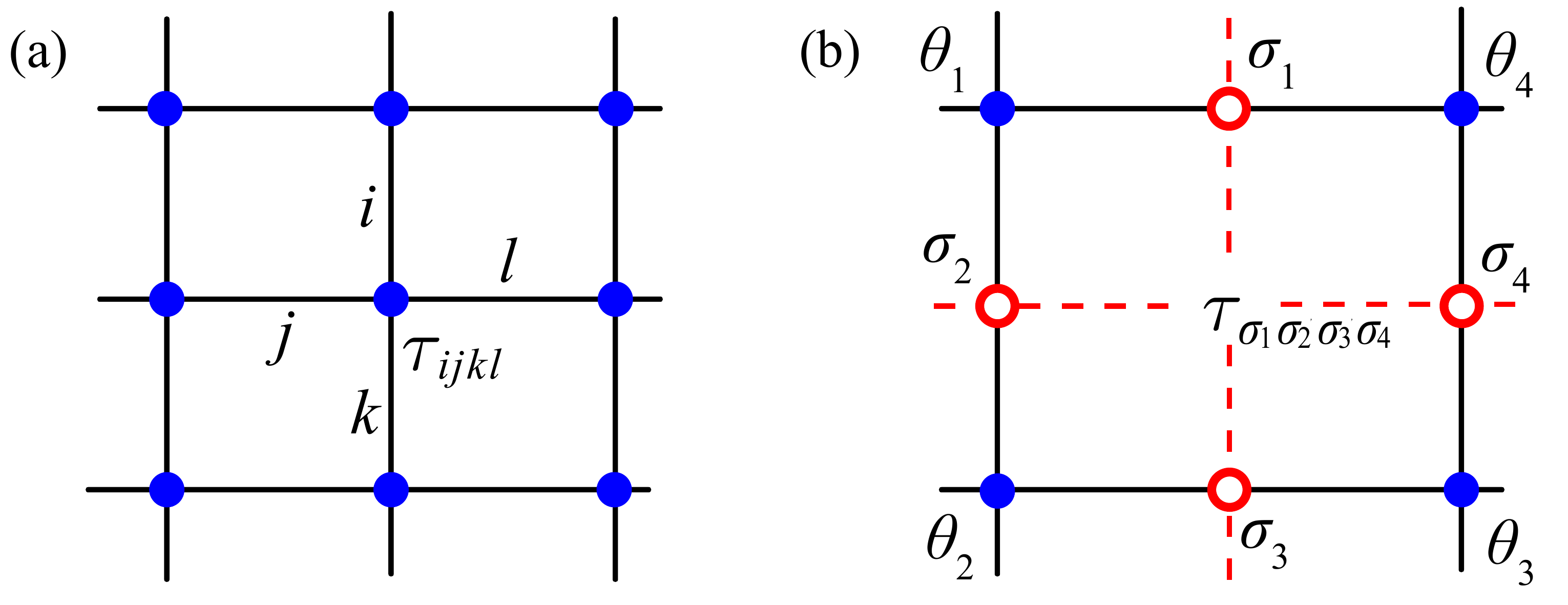}
\caption{Tensor network representation of the partition function for the $q$-state clock model in (a) the original lattice and (b) its dual lattice. The dual spin $\sigma_\alpha$ is defined on each bond and given by the difference between the original spins at the two ends of the bond. For example, $\sigma_1 = \theta_1 - \theta_4$.}
\label{Fig:lattice}
\end{figure}

{\em Tensor network method}. 
In the tensor network framework, the partition function of the $q$-state clock model is represented as a contraction of local tensors,
\begin{equation}
Z = \sum_{\{ \ldots ijkl \ldots \}} \left( \cdots \tau_{ijkl} \cdots \right) ,
\end{equation}
where the local tensor $\tau_{ijkl}$, living at each vertex of the square lattice [see Fig.~\ref{Fig:lattice}(a)], is given by~\cite{JingChen,Supple}
\begin{equation}
\tau_{ijkl} = \sqrt{\lambda_i \lambda_j \lambda_k \lambda_l}\delta_{\mathrm{mod}(i+j-k-l,q)} \, ,
\label{eq:original_T}
\end{equation}
where $\lambda_m=\sum_q \cos(m\theta_q ) \exp (\beta \cos \theta_q)/\sqrt{q}$ and each tensor index runs from $0$ to $q-1$.

Similarly, one can also express the partition function as a tensor-network in the dual lattice~\cite{JingChen,Supple}. The square lattice is self-dual and the dual spin is defined at each bond that connects two neighboring sites in the original lattice (Fig.~\ref{Fig:lattice}(b)). In the dual representation, the local tensor has exactly the same form as in Eq. (\ref{eq:original_T}), but $\lambda_i$ now changes to $\lambda_i = \sqrt{q} \exp (\beta\cos\sigma )$. The dual tensor-network model is related to the original clock model via the Kramers-Wannier dual transformation~\cite{Kramers1941}. Under this transformation, the low and high temperature phases are exchanged.

We use the  variational uniform matrix product state (VUMPS) algorithm~\cite{VUMPS} to evaluate physical quantities that are needed in order to quantify the critical behaviors of the model in the thermodynamic limit. VUMPS is to find a fixed-point MPS solution to approximate the largest eigenvector of the row-to-row transfer matrix~\cite{Supple}. The accuracy of this approximation is controlled by the bond dimension $D$ of local tensors. Other physical quantities, such as the entanglement entropy $S_{\mathrm{E}}$ and the correlation length $\xi$, can be determined from the fixed-point MPS~\cite{Supple}.

\begin{figure}[tpb]
\centering
\includegraphics[width=8.5cm]{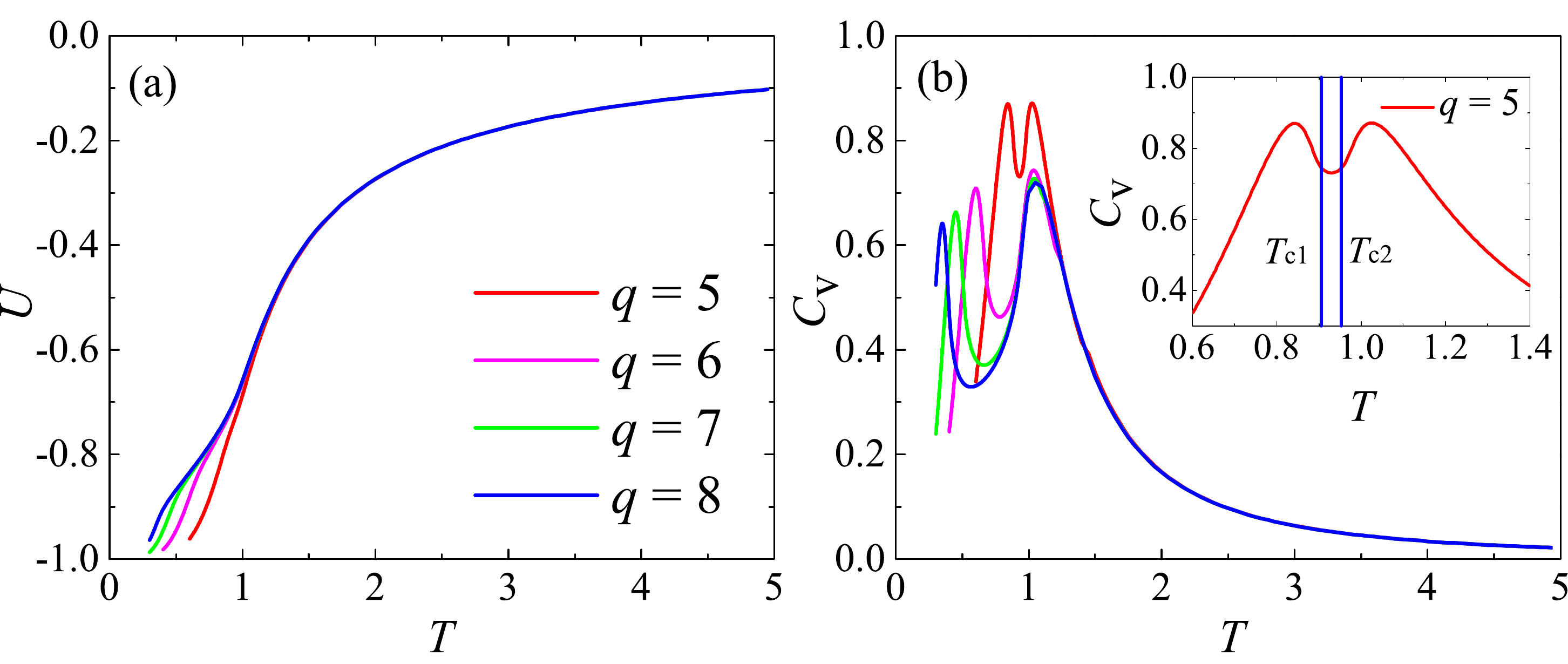}
\caption{ (a) The internal energy and (b) specific heat as a function of temperature $T$ for the $q$-state clock model obtained by VUMPS with $D = 250$.}
\label{Fig:UCv}
\end{figure}

{\em Critical properties}.
Similar as in the classic $XY$ model, we find that there is not any singularity in the temperature dependence of the internal energy, the specific heat (Fig.~\ref{Fig:UCv}) and other thermodynamic quantities for the $q$-state clock model. The peak positions of the domes in the specific heat curve do not correspond to the critical transition temperatures which are indicated by the vertical lines in the inset of Fig.~\ref{Fig:UCv}(b) for the $q=5$ clock model. This gives the first indication that the phase transitions are of the infinite-order BKT type~\cite{Berezinskii,KT1,KT2,KT3}.

\begin{figure}[tp]
\centering
\includegraphics[width=8.5cm]{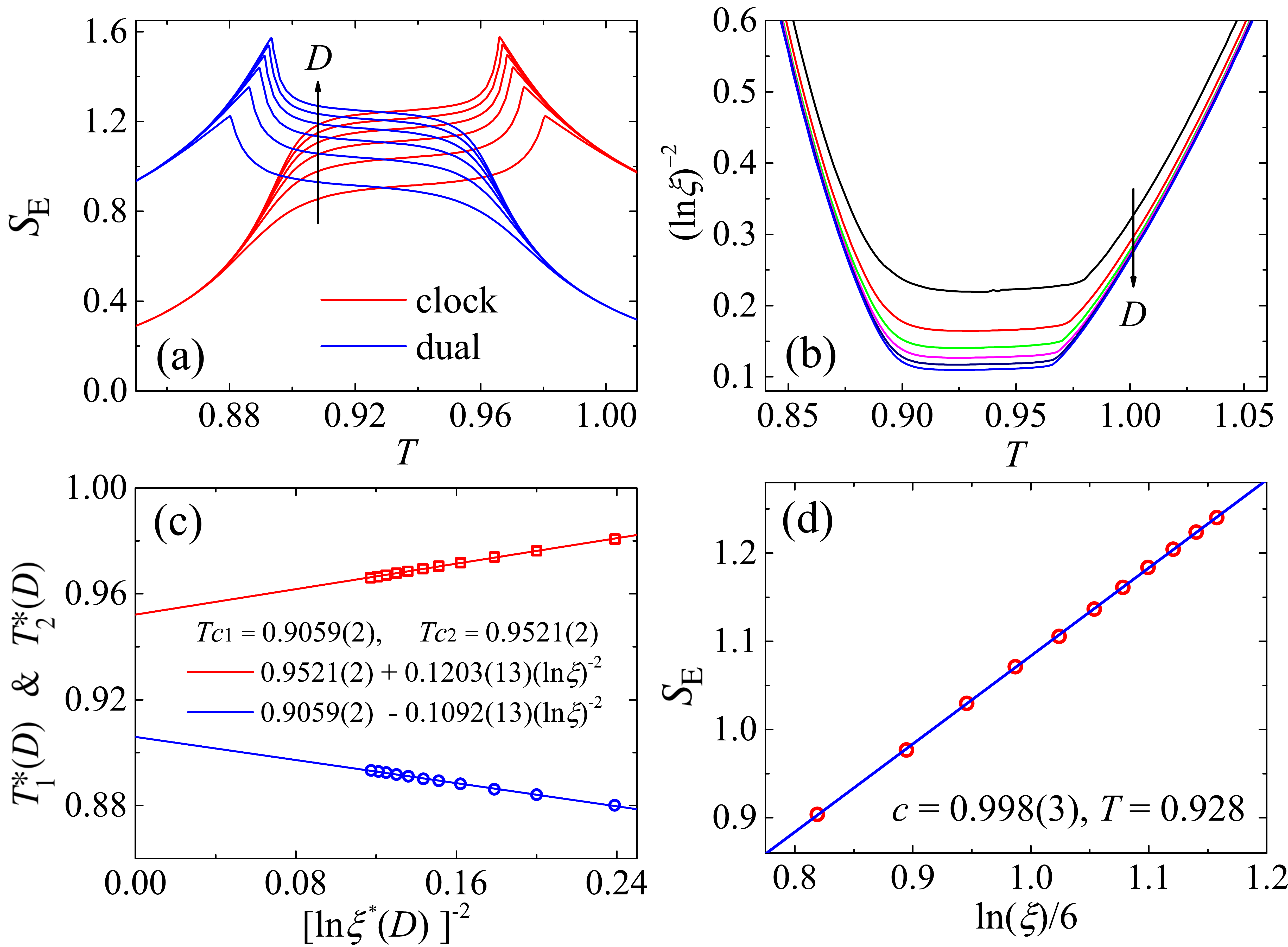}
\caption{ (a) The entanglement entropy $S_\mathrm{E}$ of the boundary MPS as a function of temperature for the $q=5$ clock model (red curves) and its dual lattice model (blue curves). (b) $(\ln\xi)^{-2}$ ($\xi$ is the correlation length) as a function of temperature for the $q=5$ clock model. The arrow indicates the increase of $D$ from $50$ to $250$ with an interval of 40. (c) The peak temperatures $T_{1}^{*}(D)$ and $T_{2}^{*}(D)$ of the entanglement entropy as a function of  $[\ln\xi^{*}(D)]^{-2}$,  where $\xi^{*}(D)$ is the correlation length at the peak position. The solid curves are linear fits to the data. (d) Central charge $c=0.998(3)$ extracted from the linear fit of $S_\mathrm{E}$ as a function of $\mathrm{ln} (\xi)$ at $T=0.928$ for the $q=5$ clock model.
}
\label{fig:q5trans}
\end{figure}

However, we find that the transitions do induce singularities in the temperature dependence of the entanglement entropy of the fixed-point MPS, $S_\mathrm{E}$. This offers an accurate approach to determine the critical transition temperatures. As shown in Fig.~\ref{fig:q5trans}(a), $S_{\text{E}}$ develops sharp peaks at two critical points, $T_{1}^{*}(D)$ and $T_{2}^{*}(D)$, for the $q=5$ clock model in the original (red curves) and its dual lattice (blue curves), respectively. By extrapolating the bond dimension $D$ to infinite, we can determine accurately the critical temperatures from $T_{1}^{*}(D)$ and $T_{2}^{*}(D)$ of $S_\mathrm{E}$.

Furthermore, we find that the inverse square of the logarithmic correlation length, $(\ln\xi)^{-2}$, varies linearly on the temperature when critical points are approached from the off-critical phases (Figure~\ref{fig:q5trans}(b)). Hence the correlation length $\xi$ scales exponentially with $T$ as
\begin{equation}
\xi(T) \propto \exp \left(b/\sqrt{|T-T_{c}|}\right) .
\label{Eq:xi_Tc}
\end{equation}
This is a characteristic and unique property of the BKT transition~\cite{KT3}.

Figure~\ref{fig:q5trans}(c) shows how the peak temperatures, $T_{1}^{*}(D)$ and $T_{2}^{*}(D)$, vary with $\ln^{-2} \xi (D)$. The linear variance of $T_{1}^{*}(D)$, similarly $T_{2}^{*}(D)$, with $\ln^{-2} \xi (D)$ agrees perfectly with Eq.~(\ref{Eq:xi_Tc}). This allows us to estimate the critical temperatures, $T_{c1}$ and $T_{c2}$, simply by linear extrapolation of $T_{1}^{*}(D)$ and $T_{2}^{*}(D)$ with respect to $\ln^{-2} \xi (D)$.

Within the critical phase, it is known that the entanglement entropy $S_\mathrm{E}$ scales logarithmically with the corresponding correlation length $\xi$,  $S_\mathrm{E} \propto (c/6) \ln\xi$~\cite{SE_scaling,Pollmann2009,Pirvu2012}, where $c$ is the central charge of conformal field theory (CFT). From a linear slope of $S_\mathrm{E}$ as a function of $\ln \xi  $ at a representative temperature $T=0.928$ within the critical phase of the 5-state clock model, we find that $c$ equals $1$ within $0.1\%$ [Fig.~\ref{fig:q5trans}(d)], in agreement with the prediction of conformal field theory.

The above analysis has also been carried out for the clock models with $q=6,7,8$ and their dual models. For all the cases we have studied, we find that the critical transitions belong to the BKT-type. A detailed discussion is given in the SM~\cite{Supple}.


{\em Comparison with the prediction of field theory}.
In the large $q$ and long wavelength limit, it is known that the $q$-state clock model is described by the so-called $\mathbb{Z}_q$-deformed sine-Gordon theory~\cite{Wiegmann1978,Matsuo2006}
\begin{eqnarray}
S&=&\frac{1}{2 \pi K} \int d^{2} \mathbf{r}(\nabla \phi)^{2}+\frac{g_{1}}{2 \pi \alpha^{2}} \int d^{2} \mathbf{r} \cos (\sqrt{2} \phi) \nonumber\\
&+& \frac{g_{2}}{2 \pi \alpha^{2}} \int d^{2} \mathbf{r} \cos (q \sqrt{2} \Theta),
\label{Eq:effectiveModel}
\end{eqnarray}
where the real scalars $\phi$ and $\Theta$, being compactified on a circle as $\phi \equiv\phi+\sqrt{2} \pi$ (similarly for $\Theta$), are mutually dual to each other, i.e., $\partial_{x} \phi=-K\partial_{y}\Theta$ and $\partial_{y} \phi = K \partial_{x} \Theta$. $\alpha$ is an ultraviolet cutoff. The coupling constants $K$, $g_1$, and $g_2$ are temperature dependent, but their functional forms are
\textit{a priori} unknown.

Without the last term, Eq. (\ref{Eq:effectiveModel}) is just the action of the standard sine-Gordon theory describing the classical $XY$ model. The O(2) spin-rotational symmetry of the $XY$ model corresponds to the invariance of the action under the transformation $\Theta \rightarrow \Theta + \gamma$. In the presence of the $q$-dependent cosine-potential,
the O(2) symmetry group reduces to $\mathbb{Z}_q$, and the action is variant only when $\gamma = \sqrt{2}\pi m/q$ ($m=0, \ldots ,q-1$). If $K=q$ and $g_1 = g_2$, the deformed sine-Gordon model $(\ref{Eq:effectiveModel})$ becomes \textit{self-dual}~\cite{Lecheminant2002,LiWei2015}, i.e., the action is invariant under the dual transformation $\phi \leftrightarrow q \Theta$, which corresponds to the Kramers-Wannier dual transformation~(\ref{Eq:lattice_dual_transformation}) on the lattice.

The phase diagram of the effective theory (\ref{Eq:effectiveModel}) can be understood from the renormalization group flow of the second and third terms under the scaling transformation. The scaling dimensions of these terms are given by
\begin{equation}
\Delta_{\cos (\sqrt{2} \phi)} =\frac{K}{2}, \qquad \Delta_{\cos (q \sqrt{2} \Theta)} =\frac{q^{2}}{2 K} .
\label{Eq:OscalingDimension}
\end{equation}
These terms are relevant or irrelevant when their scaling dimensions are smaller or larger than 2. They become marginal (with scaling dimension 2) at $K_{c2} = 4$ and $K_{c1} = q^2/4$, respectively. At the self-dual point ($K_{\mathrm{sd}} = q$), they have the same scaling dimension $\Delta = q/2$.

The value of $K$ increases with decreasing temperature. Thus the second term is relevant in high temperatures, turns marginal at a critical temperature $T_{c2}$ where $K_{c2} = 4$, and becomes irrelevant in low temperatures. This term would drive the system into a noncritical paramagnetic phase at high temperature. On the contrary, the third term is irrelevant and becomes relevant until the temperature drops below another critical point $T_{c1}$ so that $K$ becomes larger than $K_{c1}$. This term would drive the system into a ferromagnetic phase at low temperature. When $q>4$, there is a gap between $T_{c1}$ and $T_{c2}$, in which both the second and third terms are irrelevant and the system lies in a critical phase. The two critical temperatures are related to each other through the dual transformation $\phi \leftrightarrow q \Theta$. For smaller $q$ ($q=2,3,4$), the intermediate critical phase shrinks into one point with $T_{c1}=T_{c2}$ and the transition from the paramagnetic to ferromagnetic becomes conventional Landau-Ginzburg type~\cite{Lecheminant2002,LiWei2015,Supple}.

In the noncritical paramagnetic phase, $T > T_{c2}$, the effective action does not depend much on the third term since it is irrelevant. Thus the thermodynamics of the system should be approximately $q$-independent in this phase away from the critical point. This is indeed consistent with our numerical results (Fig.~\ref{Fig:UCv}). This phenomenon was first observed in Ref.~\cite{Lapilli2006} and was termed as ``extended universality''. Nevertheless, it is worth mentioning that the conclusions in Ref.~\cite{Lapilli2006} about the BKT transitions are different from ours. Instead, they claimed that the transitions at $T_{c2}$ are not of the BKT type for $q < 8$~\cite{Lapilli2006}, based on the reasoning that the helicity modulus with respect to an infinitesimal twist~\cite{Minnhagen2003} was finite and continuous at $T_{c2}$ for $q < 8$. However, after using a finite and quantized twist in the definition of helicity modulus, later works~\cite{kumano,DMRG_q5} arrived at conclusions which are in agreement with ours.

Within the critical phase, the $\mathbb{Z}_q$-deformed sine-Gordon model (\ref{Eq:effectiveModel}) is dictated by the first term. Thus the critical behavior is governed purely by parameter $K$, and the critical exponents or the scaling dimensions should depend solely on $K$. For example, the critical exponent $\eta$ of the spin-spin correlation function equals $1/K$~\cite{Supple}, and $\eta = 1/K_{c1}=4/q^2$, $1/K_{\mathrm{sd}} =1/q$ and $1/K_{c2} =1/4$ at $T =T_{c1}$,  $T_{\mathrm{sd}}$ and  $T_{c2}$, respectively.

After dropping the irrelevant terms, the effective action describing the critical phase becomes~\cite{Supple}
\begin{equation}
S \simeq \frac{1}{8 \pi} \int d^{2} \mathbf{r}\left(\nabla \Theta^{\prime}\right)^{2},
\label{Eq:Seff}
\end{equation}
where $\Theta^\prime  = 2\sqrt{K}\Theta$. This is the action of the compactified boson CFT with central charge $c=1$. The scalar field $\Theta^{\prime}$ is now compactified as $\Theta^{\prime}\equiv \Theta^{\prime} + 2\pi R$ with the compactification radius $R = \sqrt{2K} $~\cite{Supple}. For this compactified boson CFT,  it has been shown~\cite{Tang2019a} that the compactification radius $R$ is determined by the ratio between the partition function defined on the Klein bottle and that on the torus
\begin{equation}
g = \frac{Z^{\mathcal{K}}\left(2 L_x, L_y/2\right)}{Z^{\mathcal{T}}(L_x, L_y)} = R,
\label{Eq:Klein_Entropy}
\end{equation}
which is valid in the limit $L_x \gg L_y$. $L_x$ and $L_y$ are the length scales along the $x$- and $y$-axis directions, respectively. $Z^{\mathcal{K}}$ and $Z^{\mathcal{T}}$ are the partition functions defined on the Klein bottle and torus manifold, respectively. The Klein bottle imposes a special boundary condition for the field configurations. $g$ is called the Klein bottle ratio. In the effective field theory, this ratio at the BKT transition points and the self-dual temperature is predicted to be
\begin{eqnarray}
g_{c1} = R_{c1} &=& \sqrt{2K_{c1}}  = q/\sqrt{2} ,  \\
g_{c2} = R_{c2} &=& \sqrt{2K_{c2}}  = 2\sqrt{2} ,  \\
g_{\mathrm{sd}} = R_{\mathrm{sd}} &=& \sqrt{2K_{\mathrm{sd}}} = \sqrt{2q} .
\end{eqnarray}

\begin{figure}[t]
\centering
\includegraphics[width=8.5cm]{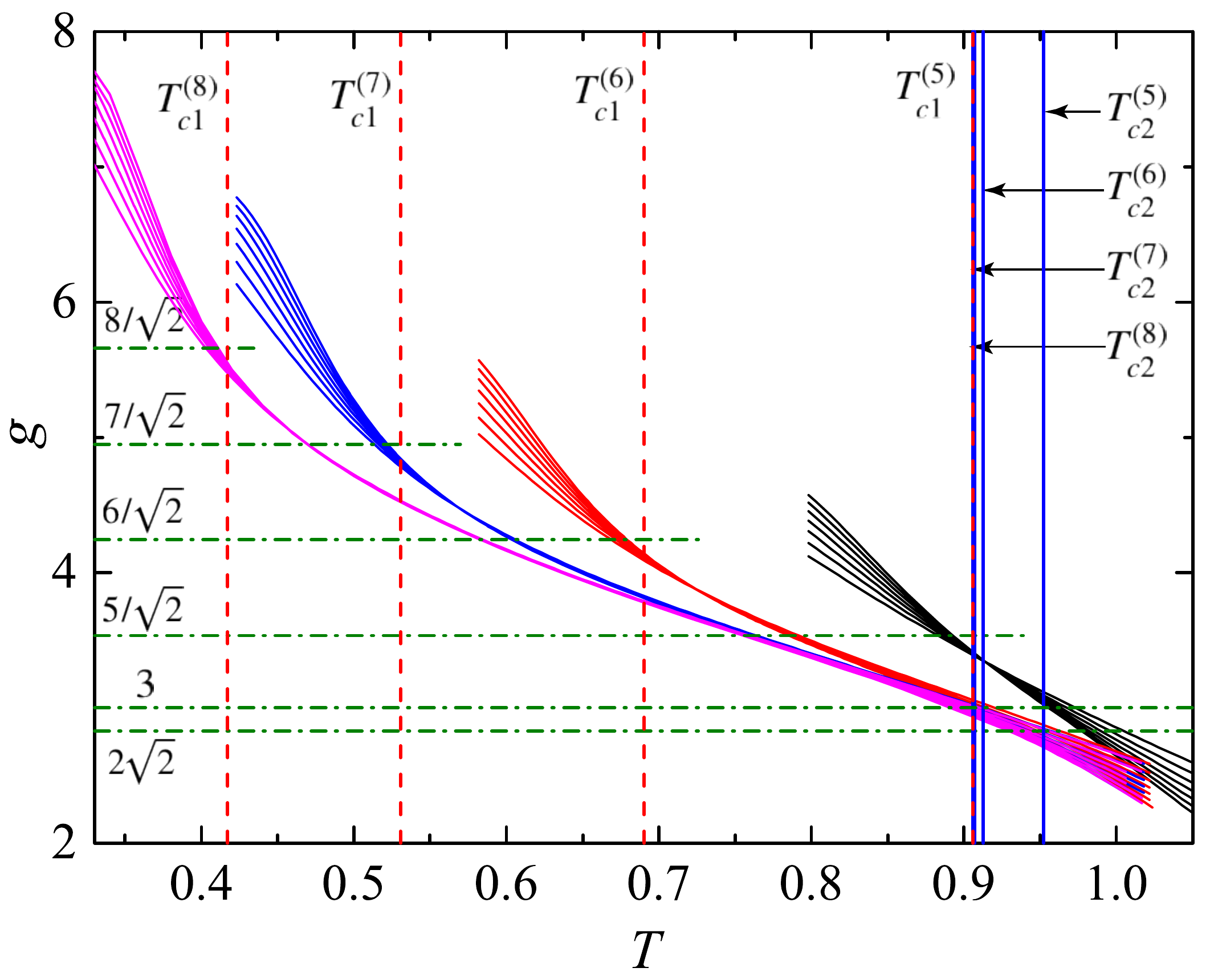}
\caption{Temperature dependence of the Klein bottle ratios $g$ for the clock models ($q = 5, 6, 7, 8$) obtained with $L_y$ running from $8$ to $20$ (different curves with the same color) and $D = 250$. The transition temperatures (vertical lines) and the predicted Klein bottle ratios (horizontal lines) are shown for reference.}
\label{Fig:KleinRatio}
\end{figure}

We have evaluated the Klein bottle ratio using the Klein bottle entropy approach~\cite{Tu2017,Tang2017a,Chen2017a,Wang2018,Tang2019a}. The Klein bottle and torus boundary conditions can be readily imposed in the tensor network framework. In the calculation, we set $L_x \rightarrow \infty$ and determine the value of $K$ by analyzing the scaling behavior of $g$ with $L_y$~\cite{Wang2018}. We use the density matrix renormalization group to evaluate the largest eigenvalues and the corresponding eigenvectors of the column-to-column transfer matrix (Fig.~\ref{Fig:Klein_approach} in the SM) and then the partition function ratio~\cite{Supple}.

For the clock model with $q = 5, 6, 7, 8$, our numerical results for the Klein bottle ratio $g$ as a function of temperature are shown in Fig.~\ref{Fig:KleinRatio}. Within the critical phase, the Klein bottle ratios for $q=6,7,8$ show good data collapse for different $L_y$, in agreement with the field theory prediction. For $q = 5$, two transition points are rather close, and the data collapse is too narrow to be observed in such a small interval. However, at both transition points, visible deviations of $g$ from the field theory prediction are observed. More specifically, the Klein bottle ratio $g$ is greater than the field theory value $2\sqrt{2}$ at $T_{c2}$, but smaller than the field theory value $q/\sqrt{2}$ at $T_{c1}$. This behavior is consistent with the discrepancy between the exponent $\eta$ obtained in the Monte Carlo simulations~\cite{Surungan2019} and that predicted by CFT $\eta=1/K$~\cite{Elitzur,KT3}. These deviations result from some marginal operators that are not included in the effective field theory (\ref{Eq:Seff}). These terms may lead logarithmic corrections to the critical exponents~\cite{KT3,Janke1997}. The sizable contributions of marginal operators to the Klein bottle ratio were also observed in other  models~\cite{Tang2017a,Tang2019a}.

\begin{figure}[t]
\centering
\includegraphics[width=8.5cm]{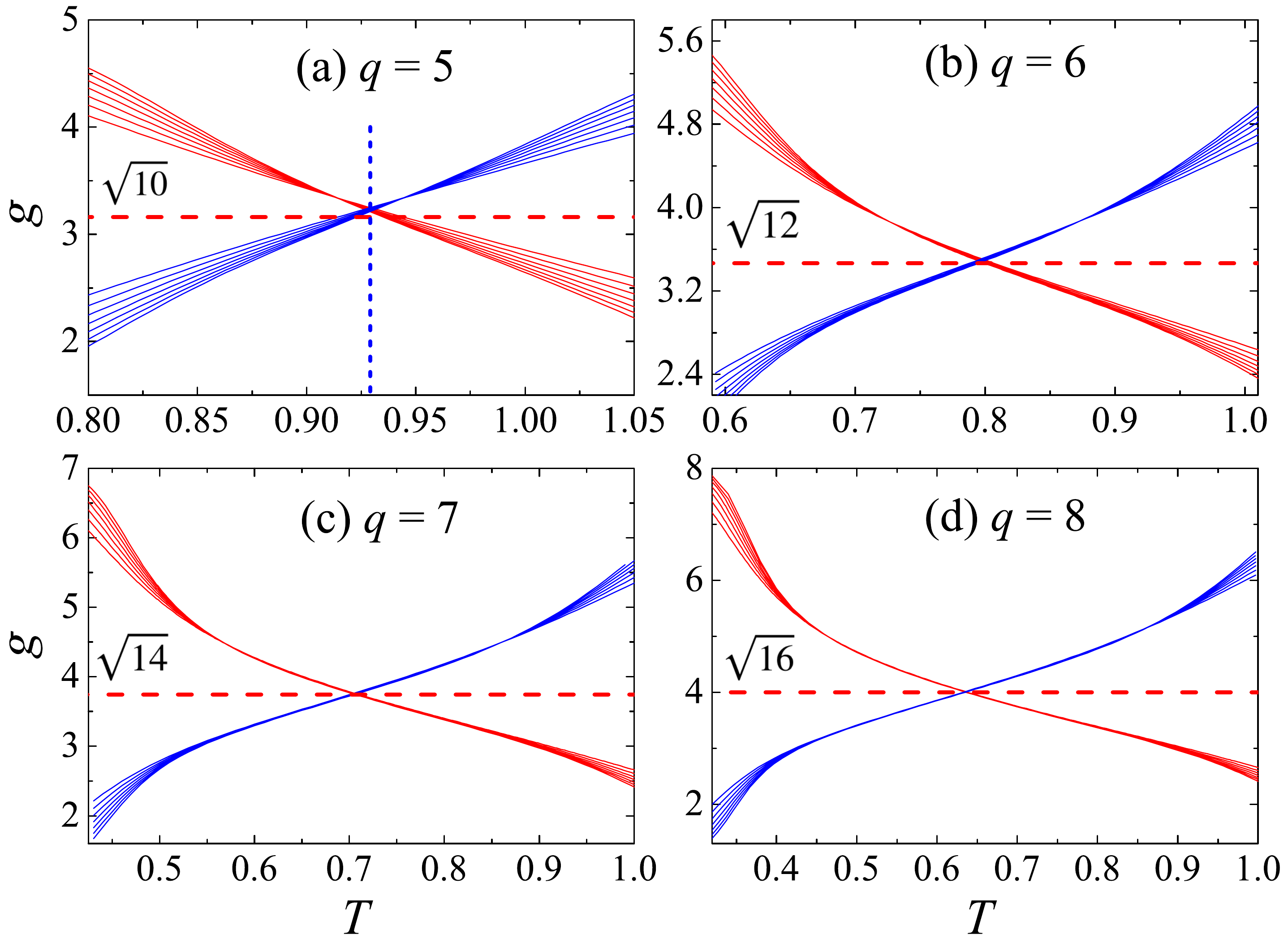}
\caption{(a)-(d) Temperature dependence of the Klein bottle ratio for the clock models with $q = 5, 6, 7, 8$ (red curves) and their dual models (blue curves). $L_y$ runs from 8 to 20. The horizontal dashed lines correspond to the field theory prediction $g = \sqrt{2q}$ for the self-dual cases. For $q=5$, the exact self-dual temperature is indicated by the vertical dashed line.}
\label{Fig:dualKsd}
\end{figure}

Finally, we comment on the possible existence of a self-dual point in the clock model. For $q = 5, 6, 7, 8$, the Klein bottle ratios $g$ for the clock models
and their dual models are plotted in Fig.~\ref{Fig:dualKsd}. For $q = 6, 7, 8$, these curves intersect approximately at a single point (such intersection occurs at
the self-dual temperature $T_{\mathrm{sd}}$ for $q=2,3,4$; see Fig.~\ref{Fig:Klein_q234} in the SM), which is the expected self-dual point, and the Klein bottle ratios at this point coincide with the predicted value $g_{\mathrm{sd}} = \sqrt{2q}$. Furthermore, we find that the free energy and other thermodynamic quantities of the clock model at rescaled temperature $T/T_{\mathrm{sd}}$  coincide with good precision with the corresponding quantities of the dual lattice model at a rescaled inverse temperature $T_{\mathrm{sd}}/T$ (see Fig.~\ref{Fig:DualMSX} in the SM). The existence of the self-dual point in the large-$q$ clock model suggests that the effective field theory for describing the clock model should be a self-dual $\mathbb{Z}_q$-deformed sine-Gordon model with $g_1 = g_2$.

{\em Conclusion}.
In summary, we have clarified the nature of critical phases and phase transitions in the two-dimensional $q$-state clock model by
combining state-of-the-art tensor network simulations with field-theory analysis. We have provided extensive evidence to show that there is an intermediate critical phase of $c=1$ and the critical transitions at  $T_{c1}$ and $T_{c2}$ are of the BKT type for all the cases with $q>4$. For these infinite-order transitions, we find that a singular peak develops in the temperature dependence of the entanglement entropy of the fixed-point MPS, although there is not any singularity in the conventional thermodynamic quantities. This offers a unique opportunity for us to determine accurately the critical temperatures from scaling behavior of these singular peaks. Furthermore, we show that the long wavelength $q$-state clock model can be well described by the effective self-dual $\mathbb{Z}_q$-deformed sine-Gordon theory, and an ``extended universality'' with $q$-independent thermodynamics exists in the paramagnetic phase~\cite{Lapilli2006}. We have also calculated the compactification radius $R$ that governs the universal scaling behaviors in the critical phase.

{\em Acknowledgments}. 
We are grateful to Lei Wang and Rui-Zhen Huang for useful discussions. The authors are supported by the National Key Research and Development Project of China (Grant No.~2017YFA0302901), the National Natural Science Foundation of China (Grants No.~11888101 and No.~11874095), the Strategic Priority Research Program of Chinese Academy of Sciences (Grant No.~XDB33020300) and the DFG through project A06 of SFB 1143 (Project ID~247310070).

\bibliography{qclock_arXiv}

\clearpage

\setcounter{table}{0}
\setcounter{figure}{0}
\setcounter{equation}{0}

\renewcommand\thefigure{S\arabic{figure}}
\renewcommand\theequation{S\arabic{equation}}
\renewcommand{\thetable}{S\arabic{table}}

\begin{widetext}
\section*{Supplemental Material for "Critical properties of the two-dimensional $q$-state clock model"}

\section*{A. Tensor network representation of the $q$-state clock model and its dual model}
\label{section_SI}
The ferromagnetic $q$-state clock model on the 2D square lattice is defined by the Hamiltonian
\begin{equation}
H = -J\sum_{\left\langle ij \right\rangle}\cos(\theta_i-\theta_j),
\label{SEq:qclock_Model}
\end{equation}
where ${\left\langle ij \right\rangle}$ refers to the nearest neighbors, $\theta_{i}= 2 \pi k/q$ denotes a discrete angle variable
with $k=0,1, \ldots, q-1$ at site $i$, and $J=1$ sets the energy scale.

To derive the tensor network representation of the partition function, we use the character expansion of the local Boltzmann weight
\begin{equation}
e^{\beta \cos(\theta_i-\theta_j)} = \sum_{m=0}^{q-1} V_{\theta_i,m}\lambda_m V^{*}_{\theta_j,m},
\end{equation}
where $V_{\theta_i,m} = e^{im\theta_i}/\sqrt{q}$ is a unitary matrix, and $\lambda_m$ is a diagonal matrix (bond spectrum) defined by
\begin{equation}
\lambda_m=\sum_{\theta}{e^{-im\theta}e^{\beta \cos(\theta)}/\sqrt{q}}
=\sum_{\theta}{\cos(m\theta)e^{\beta \cos(\theta)}/\sqrt{q}}.
\label{eq:lambda}
\end{equation}
Then, the partition function $Z=\sum_{\{ \theta_i \}}e^{-\beta H}$ can be cast into the following tensor network representation:
\begin{equation}
\label{eq:partfnc}
    \begin{tikzpicture}[every node/.style={scale=0.9},scale=.45]
      \draw[scale=1.0] (-5.0,3.0) node (X) {$Z = \mathrm{Tr}$};
      \lineV{-1.0}{7.0}{-3.0}
      \lineH{-3.0}{-2.75}{7.0}
      \lineH{-3.0}{-2.75}{-1.0}
      \draw (-2.0,0.0) node (X) {$\dots$};
      \MPOtens{0.0}{0.0}{\tau}
      \MPOtens{2.0}{0.0}{\tau}
      \MPOtens{4.0}{0.0}{\tau}
      \MPOtens{6.0}{0.0}{\tau}
      \draw (8.0,0.0) node (X) {$\dots$};
      \draw (-2.0,2.0) node (X) {$\dots$};
      \MPOtens{0.0}{2.0}{\tau}
      \MPOtens{2.0}{2.0}{\tau}
      \MPOtens{4.0}{2.0}{\tau}
      \MPOtens{6.0}{2.0}{\tau}
      \draw (8.0,2.0) node (X) {$\dots$};
      \draw (-2.0,4.0) node (X) {$\dots$};
      \MPOtens{0.0}{4.0}{\tau}
      \MPOtens{2.0}{4.0}{\tau}
      \MPOtens{4.0}{4.0}{\tau}
      \MPOtens{6.0}{4.0}{\tau}
      \draw (8.0,4.0) node (X) {$\dots$};
      \draw (-2.0,6.0) node (X) {$\dots$};
      \MPOtens{0.0}{6.0}{\tau}
      \MPOtens{2.0}{6.0}{\tau}
      \MPOtens{4.0}{6.0}{\tau}
      \MPOtens{6.0}{6.0}{\tau}
      \draw (8.0,6.0) node (X) {$\dots$};
      \lineV{-1.0}{7.0}{9.0}
      \lineH{8.75}{9.0}{7.0}
      \lineH{8.75}{9.0}{-1.0}
  \end{tikzpicture}
\end{equation}
where the local tensor $\tau_{ijkl}$ [see Fig.~\ref{Fig:lattice}(a)] at each site is given by~\cite{JingChen}
\begin{equation}
\tau_{ijkl} = \sqrt{\lambda_i \lambda_j \lambda_k \lambda_l}\delta_{\mathrm{mod}(i+j-k-l,q),0}.
\label{eq:original_T}
\end{equation}


Furthermore, the dual spin variables can be defined with a Kramer-Wannier dual transformation~\cite{Kramers1941} as follows [see Fig.~\ref{Fig:lattice}(b)]:
\begin{equation}
\begin{array}{c}
\sigma_{1}=\theta_{1}-\theta_{4}, \qquad \sigma_{2}=\theta_{2}-\theta_{1},  \qquad \sigma_{3}=\theta_{2}-\theta_{3},  \qquad \sigma_{4}=\theta_{3}-\theta_{4}.
\end{array}
\label{Eq:lattice_dual_transformation}
\end{equation}
Thus, the partition function of the dual model can also be represented as a tensor network, defined by another local tensor
\begin{equation}
\tau_{\sigma_{1}\sigma_{2}\sigma_{3}\sigma_{4}} = \sqrt{\Lambda_{\sigma_{1}} \Lambda_{\sigma_{2}} \Lambda_{\sigma_{3}}
\Lambda_{\sigma_{4}}}  \delta_{\mathrm{mod}(\sigma_{1}+\sigma_{2}-\sigma_{3}-\sigma_{4},q),0},
\label{eq:dual_tensor}
\end{equation}
where $\Lambda_{\sigma} = \sqrt{q} e^{\beta\cos\sigma}$.

One advantage of the tensor network representation is that the self-dual points for $q=2,3,4,5$ become manifest
\begin{equation}
\beta_{\mathrm{sd}}=\left\{\begin{array}{ll}{\frac{1}{2} \ln (\sqrt{2}+1),} & {q=2} \\ {\frac{2}{3} \ln (\sqrt{3}+1),} & {q=3} \\ {\ln (\sqrt{2}+1),} & {q=4}\end{array}\right.
\end{equation}
and
\begin{equation}
\frac{e^{5 \beta_{\mathrm{sd}} / 4}}{\cosh \left(\sqrt{5} \beta_{\mathrm{sd}} / 4\right)}=\sqrt{5}+1,  \qquad q = 5.
\label{eq:q5betac}
\end{equation}
For $q = 5$, the numerical solution to Eq.~(\ref{eq:q5betac}) gives $\beta _{\mathrm{sd}} \approx 1.0763180716046478$ ($T_{\mathrm{sd}} \approx
0.929093384550472$)~\cite{JingChen}.

\section*{B. VUMPS algorithm and results}

This section provides a brief review of the VUMPS algorithm~\cite{VUMPS}. We emphasize on how to use it to calculate the
physical quantities, such as the free energy, magnetization, internal energy, entanglement entropy, and correlation length.

The VUMPS method works in the thermodynamic limit and determines the fixed-point MPS, which approximates the leading eigenvector
corresponding to the largest eigenvalue of the row-to-row transfer matrix, namely,
\begin{equation}
  \begin{tikzpicture}[every node/.style={scale=0.80},scale=.45]
      \MPSuLtens{0.0}{2.0}{A_L}
      \MPSuLtens{2.0}{2.0}{A_L}
      \MPSuCtens{4.0}{2.0}{A_C}
      \MPSuRtens{6.0}{2.0}{A_R}
      \MPSuRtens{8.0}{2.0}{A_R}
      \draw (-2.0,0.0) node (X) {$\dots$};
      \MPOtens{0.0}{0.0}{\tau}
      \MPOtens{2.0}{0.0}{\tau}
      \MPOtens{4.0}{0.0}{\tau}
      \MPOtens{6.0}{0.0}{\tau}
      \MPOtens{8.0}{0.0}{\tau}
      \draw (10.0,0.0) node (X) {$\dots$};
      \draw (11.0,0.0) node (X) {$\approx$};
      \draw (-3.5,-3.0) node (X) {$\lambda_{\mathrm{max}}^N$};
      \draw (-2.0,-3.0) node (X) {$\dots$};
      \MPSuLtens{0.0}{-3.0}{A_L}
      \MPSuLtens{2.0}{-3.0}{A_L}
      \MPSuCtens{4.0}{-3.0}{A_C}
      \MPSuRtens{6.0}{-3.0}{A_R}
      \MPSuRtens{8.0}{-3.0}{A_R}
      \draw (10.0,-3.0) node (X) {$\dots$};
  \end{tikzpicture}
  \label{eq:vumps}
\end{equation}
where $\lambda_{\mathrm{max}}$ is the largest eigenvalue per site of the row-to-row transfer matrix, which is related to free
energy density per site by $f = -k_{B}T \,\mathrm{ln}(\lambda_{\mathrm{max}})$.
Furthermore, the fixed-point MPS satisfies the mixed canonical form conditions
\begin{equation}
  \begin{tikzpicture}[every node/.style={scale=0.80},scale=.45]
    \cornerIultens{-3.0}{1.0}
    \cornerIdltens{-3.0}{-1.0}
    \MPSuLtens{-1.0}{1.0}{A_L}
    \MPSdLtens{-1.0}{-1.0}{\bar{A}_L}
    \draw (1.0,0.0) node {$=$};
    \cornerIultens{2.0}{1.0}
    \cornerIdltens{2.0}{-1.0}
    \draw (4.0,0.0) node {,};
    \MPSuRtens{6.0}{1.0}{A_R}
    \MPSdRtens{6.0}{-1.0}{\bar{A}_R}
    \cornerIurtens{8.0}{1.0}
    \cornerIdrtens{8.0}{-1.0}
    \draw (9.0,0.0) node {$=$};
    \cornerIurtens{11.0}{1.0}
    \cornerIdrtens{11.0}{-1.0}
   \end{tikzpicture}
\end{equation}
and $A_L$, $A_R$, and $A_C$ near or at the fixed-point satisfy
\begin{equation}
  \begin{tikzpicture}[every node/.style={scale=0.80},scale=.45]
   \MPSuLtens{-2.0}{0.0}{A_L}
   \centerCudtens{0.0}{0.0}{C}
    \draw (2.0,0.0) node {$\approx$};
   \MPSuCtens{4.0}{0.0}{A_C}
    \draw (6.0,0.0) node {$\approx$};
   \centerCudtens{8.0}{0.0}{C}
   \MPSuRtens{10.0}{0.0}{A_R}
   \end{tikzpicture}
   \label{eq:fixed_point}
\end{equation}

To determine the fixed-point tensors $A_L$, $A_R$, and $A_C$, the VUMPS algorithm includes the following four steps:
\begin{enumerate}
\item Find the left and right environments $E_L$ and $E_R$ by power or Arnoldi method:
\begin{equation}
  \begin{tikzpicture}[every node/.style={scale=0.80},scale=.45]
      \cornerIultens{0.0}{2.0}
      \MPSltens{0.0}{0.0}{E_L}
      \cornerIdltens{0.0}{-2.0}
      \MPOtens{2.0}{0.0}{\tau}
      \MPSuLtens{2.0}{2.0}{A_L}
      \MPSdLtens{2.0}{-2.0}{\bar{A}_L}
      \draw (4.0,0.0) node (X) {$\approx$};
      \draw (5.0,0.0) node (X) {$\Omega_L$};
      \cornerIultens{6.0}{2.0}
      \MPSltens{6.0}{0.0}{E_L}
      \cornerIdltens{6.0}{-2.0}
      \draw (8.0,0.0) node (X) {$,$};
      \MPOtens{10.0}{0.0}{\tau}
      \MPSuRtens{10.0}{2.0}{A_R}
      \MPSdRtens{10.0}{-2.0}{\bar{A}_R}
      \cornerIurtens{12.0}{2.0}
      \MPSrtens{12.0}{0.0}{E_R}
      \cornerIdrtens{12.0}{-2.0}
      \draw (13.0,0.0) node (X) {$\approx$};
      \draw (14.0,0.0) node (X) {$\Omega_R$};
      \cornerIurtens{16.0}{2.0}
      \MPSrtens{16.0}{0.0}{E_R}
      \cornerIdrtens{16.0}{-2.0}
  \end{tikzpicture}
\end{equation}

\item Find the central tensors $C$ and $A_C$ by power or Arnoldi method:
\begin{equation}
  \begin{tikzpicture}[every node/.style={scale=0.80},scale=.45]
    \cornerIultens{0.0}{0.0}
    \centerCudtens{2.0}{0.0}{C}
    \cornerIurtens{4.0}{0.0}
    \MPSltens{0.0}{-2.0}{E_L}
    \centerIudtens{2.0}{-2.0}
    \MPSrtens{4.0}{-2.0}{E_R}
    \draw (5.0,-1.5) node (X) {$\approx$};
    \draw (6.0,-1.5) node (X) {$\Omega_C$};
    \cornerIultens{7.0}{-1.0}
    \centerCudtens{9.0}{-1.0}{C}
    \cornerIurtens{11.0}{-1.0}
     \cornerIultens{0.0}{-4.0}
    \MPSuCtens{2.0}{-4.0}{A_C}
    \cornerIurtens{4.0}{-4.0}
    \MPSltens{0.0}{-6.0}{E_L}
    \MPOtens{2.0}{-6.0}{\tau}
    \MPSrtens{4.0}{-6.0}{E_R}
    \draw (5.0,-5.5) node (X) {$\approx$};
    \draw (6.0,-5.5) node (X) {$\Omega_{A_C}$};
    \cornerIultens{7.0}{-5.0}
    \MPSuCtens{9.0}{-5.0}{A_C}
    \cornerIurtens{11.0}{-5.0}
 \end{tikzpicture}
\end{equation}
where $\Omega_{A_C}/\Omega_C \approx \Omega_{L} \approx \Omega_{R}$ is approximately equal to the
largest eigenvalue $\lambda_{\mathrm{max}}$ of the row-to-row transfer matrix near or at the fixed point.

\item Obtain new $A_L$ and $A_R$ from $A_C$ and $C$ using polar decompositions or QR
decompositions~\cite{VUMPS} (here we use  QR decompositions)
\begin{equation}
  \begin{tikzpicture}[every node/.style={scale=0.80},scale=.45]
   \MPSuCtens{-2.0}{1.0}{A_C}
   \draw (-0.5,1.0) node {$=$};
   \MPSuLtens{1.0}{1.0}{Q_L}
   \centerCudtens{3.0}{1.0}{R_{A_C}}
   \draw (4.5,1.0) node {$=$};
   \centerCudtens{6.0}{1.0}{L_{A_C}}
   \MPSuRtens{8.0}{1.0}{Q_R}
   \centerCudtens{-2.0}{-1.0}{C}
   \draw (-0.5,-1.0) node {$=$};
   \centerCudtens{1.0}{-1.0}{Q_{Cl}}
   \centerCudtens{3.0}{-1.0}{R_C}
   \draw (4.5,-1.0) node {$=$};
   \centerCudtens{6.0}{-1.0}{L_C}
   \centerCudtens{8.0}{-1.0}{Q_{Cr}}
   \end{tikzpicture}
\end{equation}
where $Q_L, Q_R, Q_{Cl}$ and $Q_{Cr}$ are unitary matrices, and the diagonal entries of
$R_{A_C}, L_{A_C}, R_C$ and $L_C$ are positive. Then, we obtain
\begin{equation}
  \begin{tikzpicture}[every node/.style={scale=0.80},scale=.45]
   \MPSuLtens{-2.0}{0.0}{A_L}
   \draw (-0.5,0.0) node {$=$};
   \MPSuLtens{1.0}{0.0}{Q_L}
   \centerCudtens{3.0}{0.0}{Q_{Cl}^{\dagger}}
   \draw (4.5,0.0) node {$,$};
   \MPSuRtens{6.0}{0.0}{A_R}
   \draw (7.5,0.0) node {$=$};
   \centerCudtens{9.0}{0.0}{Q_{Cr}^{\dagger}}
   \MPSuRtens{11.0}{0.0}{Q_R}
   \end{tikzpicture}
\end{equation}

\item Repeat the steps 1-3, until the error of the fixed point [Eq.~(\ref{eq:fixed_point})] is less than a certain threshold.
\end{enumerate}

Once the fixed-point MPS is obtained, various physical quantities can be calculated.
The entanglement spectrum of the fixed-point MPS corresponds to the squared singular values $s_n$ of the tensor $C$ [see Eq.~(\ref{eq:fixed_point})].
Thus, the entanglement entropy $S_\mathrm{E}$ of the fixed-point MPS can be readily calculated as follows:
\begin{equation}
S_\mathrm{E} = -\sum_{n=1}^{D}s^2_n \ln s^2_n,
\label{eq:entanglement_entropy}
\end{equation}
where $D$ is the bond dimension of the fixed-point MPS, and $\sum_{n=1}^{D} s^2_n =1$.

The correlation length can be calculated by
\begin{equation}
\xi \equiv -1/\ln|\epsilon_2/\epsilon_1|,
\end{equation}
where $\epsilon_1$ and $\epsilon_2$ are the largest and the second largest eigenvalues of the transfer matrix
\begin{equation}
  \begin{tikzpicture}[every node/.style={scale=0.80},scale=.45]
    \MPSuCtens{0.0}{1.0}{A_C}
    \MPSdCtens{0.0}{-1.0}{\bar{A}_C}
   \end{tikzpicture}
\end{equation}
respectively.

The expectation value of a local physical quantity $X$, such as magnetization, can be calculated by means of an impurity tensor $\tau_{X}$
\begin{equation}
    \begin{tikzpicture}[every node/.style={scale=0.90},scale=.45]
      \draw (-4,0.0) node (X) {$\langle X \rangle$};
      \draw (-3.0,0.0) node (X) {$=$};
      \cornerIultens{-2.0}{2.0}
      \MPSltens{-2.0}{0.0}{E_L}
      \cornerIdltens{-2.0}{-2.0}
      \MPOtens{0.0}{0.0}{\tau_X}
      \MPSuCtens{0.0}{2.0}{A_C}
      \MPSdCtens{0.0}{-2.0}{\bar{A}_C}
      \cornerIurtens{2.0}{2.0}
      \MPSrtens{2.0}{0.0}{E_R}
      \cornerIdrtens{2.0}{-2.0}
        \draw[scale=1.0] (3.0,0.0) node (X) {$/$};
      \cornerIultens{4.0}{2.0}
      \MPSltens{4.0}{0.0}{E_L}
      \cornerIdltens{4.0}{-2.0}
      \MPOtens{6.0}{0.0}{\tau}
      \MPSuCtens{6.0}{2.0}{A_C}
      \MPSdCtens{6.0}{-2.0}{\bar{A}_C}
      \cornerIurtens{8.0}{2.0}
      \MPSrtens{8.0}{0.0}{E_R}
      \cornerIdrtens{8.0}{-2.0}  \end{tikzpicture}
\end{equation}
For example, the impurity tensor of the magnetization $M \equiv \langle e^{i\theta}\rangle$ is defined as
\begin{equation}
\tau_M(i,j,k,l) = \sqrt{\lambda_i \lambda_j \lambda_k \lambda_l} \, \delta_{\mathrm{mod}\left(i+j-k-l+1,q\right),0} .
\label{eq:Mag}
\end{equation}
The local physical quantities of the dual model can be calculated in a similar fashion.

Fig.~\ref{Fig:rawMSX} shows the magnetization $M$, the entanglement entropy $S_\mathrm{E}$ of the fixed-point MPS, and the
correlation length $\xi$ for $q=5,6,7$ and $8$ clock models (red curves) and their corresponding dual models (blue curves).
It is observed in Fig.~\ref{Fig:DualMSX} that the above physical quantities of the clock model at the rescaled temperature
$T/T_{\mathrm{sd}}$ well coincide with those of the dual model at the rescaled inverse temperature $T_{\mathrm{sd}}/T$,
especially for $q=5$ case, whose exact self-dual point $T_{\mathrm{sd}}\approx 0.929093384550472$ is known.
This agreement gives a strong hint on the possible existence of the self-duality.

\begin{figure*}[htb]
\centering
\includegraphics[width=17.5cm]{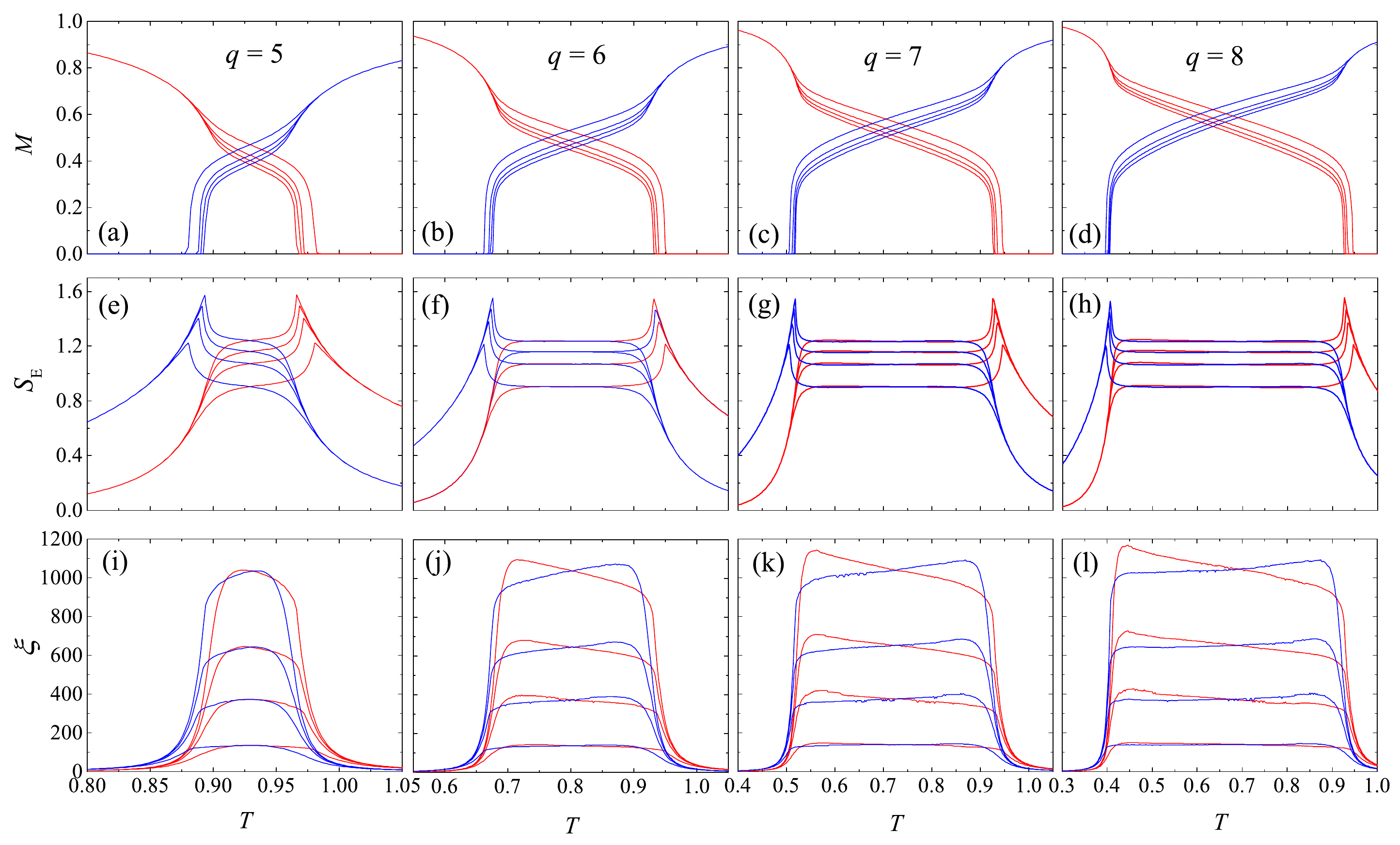}
\caption{The first row shows the magnetization as a function of temperature $T$ for the clock models (red curves) and their corresponding
dual models (blue curves) with $q=5,6,7,8$, respectively. Similarly, the second and third rows show the
entanglement entropy and the correlation length of the clock models and their dual models, respectively.
In each figure, the bond dimension for each curve from bottom to top is $D=50,110,170$ and $250$, respectively.}
\label{Fig:rawMSX}
\end{figure*}

\begin{figure*}[htb]
\centering
\includegraphics[width=17.5cm]{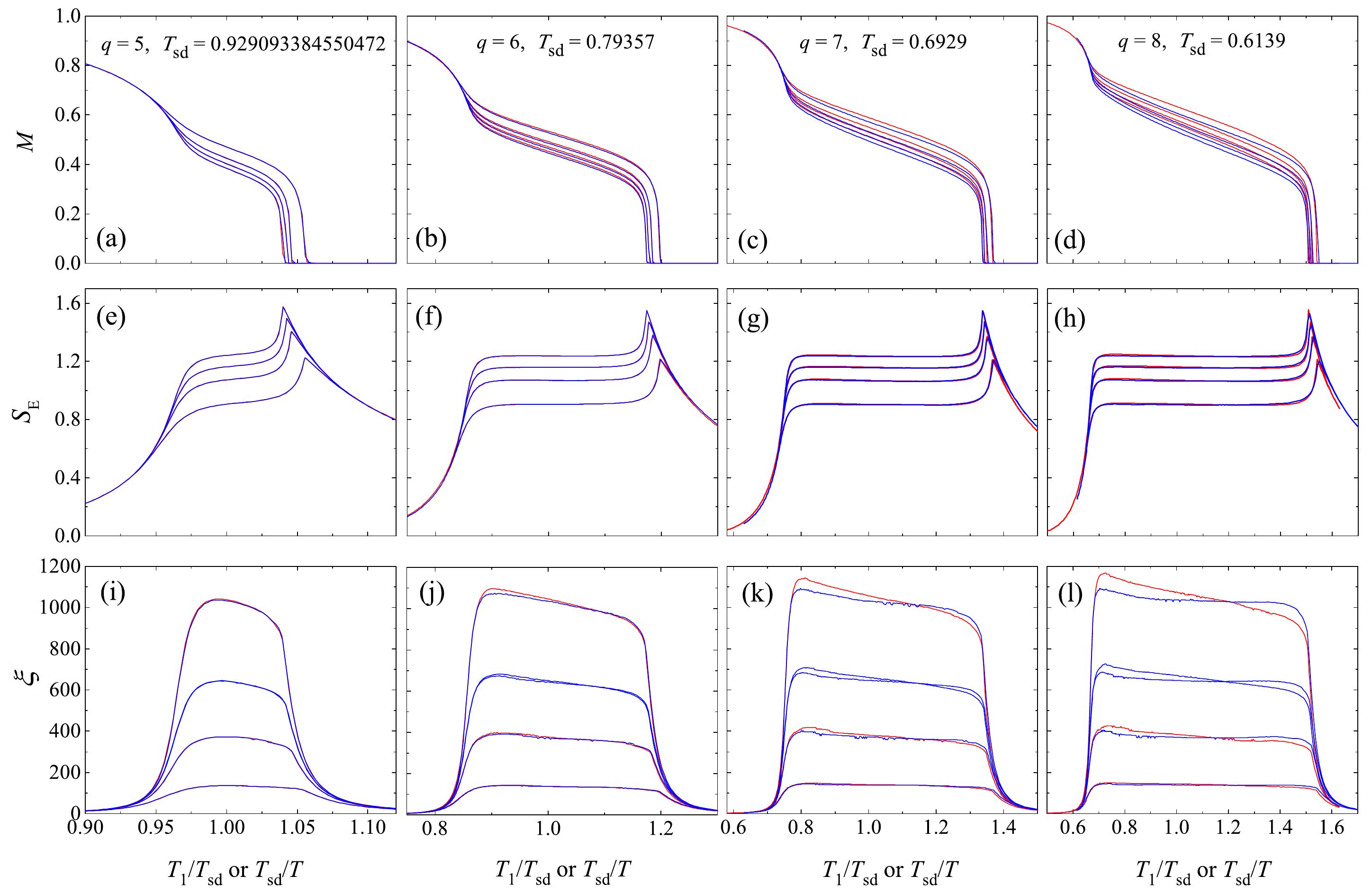}
\caption{ Magnetization $M$ (first row), entanglement entropy $S_E$ (second row) and correlation length $\xi$ (third row) as a function of the
rescaled temperature $T/T_{sd}$ for the clock model and the rescaled inverse temperature $T_{sd}/T$ for the dual model. In each figure,
the bond dimension for each curve from bottom to top is $D=50,110,170$ and $250$, respectively.}
\label{Fig:DualMSX}
\end{figure*}

\begin{figure*}[htb]
\centering
\includegraphics[width=17.5cm]{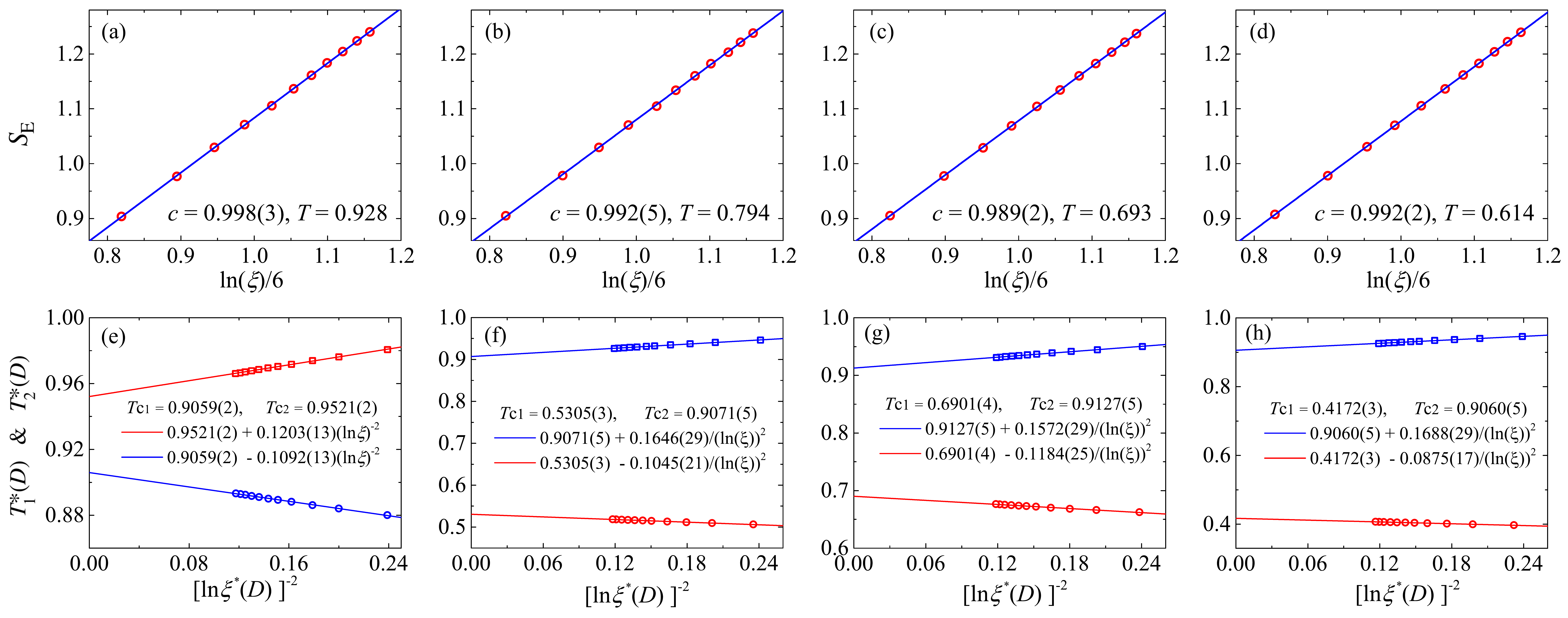}
\caption{The first row shows the central charge $c$ extracted by linearly fitting the slope between $S_\mathrm{E}$ and $\ln(\xi)/6$
at representative point $T = 0.928, 0.794, 0.693$ and $0.614$ in the critical phase for $q = 5, 6, 7$ and $8$, respectively.
The second row shows the critical temperatures $T_{c1}$ and $T_{c2}$ extracted by linearly extrapolating the peak positions
$T_{1}^{*}(D)$ and $T_{2}^{*}(D)$ of $S_\mathrm{E}$ with $(\ln\xi^{*}(D))^{-2}$, where $\xi^{*}(D)$ is the correlation length at
the peak position for a given bond dimension $D$.
}
\label{Fig:Tc_c}
\end{figure*}

The first row of Fig.~\ref{Fig:Tc_c} shows the central charge $c$ extracted by linearly fitting the slope between
$S_\mathrm{E}$ and $\ln(\xi)/6$ at representative point $T = 0.928, 0.794, 0.693$ and $0.614$ in the critical phase for the clock model with
$q = 5, 6, 7$ and $8$, respectively. The second row of Fig.~\ref{Fig:Tc_c} shows the critical temperatures $T_{c1}$ and $T_{c2}$
extracted by linearly extrapolating the peak positions $T_{1}^{*}(D)$ and $T_{2}^{*}(D)$ of $S_\mathrm{E}$ with $(\ln\xi^{*}(D))^{-2}$,
where $\xi^{*}(D)$ is the correlation length at the peak position for a given bond dimension $D$. Table~\ref{Table:Tc_List} summarizes
the critical temperatures $T_{c1}$ and $T_{c2}$ that we have obtained and those obtained in the literature.

\begin{table}[ht]
\caption{Comparison of the critical temperatures $T_{c1}$ and $T_{c2}$ for the clock model with $q = 5, 6, 7$ and $8$. }
\begin{tabular}{llllllllllllll}
\hline\hline
$q=5$                                                 &&&&   $T_{c1}$         &&&  $T_{c2}$      \\
\hline
Ref.~\cite{Tobochnik1982}(1982)       &&&&     0.8                &&&   1.1               \\
Ref.~\cite{Borisenko2011}(2011)        &&&&     0.905(1)       &&&   0.951(1)       \\
Ref.~\cite{kumano}(2013)                   &&&&     0.908           &&&   0.944           \\
Ref.~\cite{DMRG_q5}(2014)               &&&&     0.914(12)     &&&   0.945(17)    \\
Ref.~\cite{Chatterjee2018}(2018)        &&&&     0.897(1)      &&&   -                   \\
Ref.~\cite{yujifeng2018}(2018)            &&&&     0.9029(1)    &&&   0.9520(1)     \\
Ref.~\cite{Surungan2019}(2019)         &&&&     0.911(5)      &&&   0.940(5)       \\
Ref.~\cite{Hong2019}(2019)                &&&&     0.908           &&&   0.945           \\
Present work                                       &&&&      0.9059(2)    &&&   0.9521(2)     \\
\hline\hline
$q=6$                                  		 &&&&    $T_{c1}$       &&&   $T_{c2}$      \\
\hline
Ref.~\cite{Tobochnik1982}(1982)        &&&&     0.6               &&&   1.3               \\
Ref.~\cite{challa}(1986)                       &&&&     0.68(2)        &&&    0.92(1)        \\
Ref.~\cite{yamagata}(1991)                &&&&     0.68             &&&    0.90            \\
Ref.~\cite{tomita}(2002)                      &&&&     0.7014(11)   &&&    0.9008(6)    \\
Ref.~\cite{hwang}(2009)                     &&&&     0.632(2)       &&&    0.997(2)      \\
Ref.~\cite{brito}(2010)                        &&&&     0.68(1)         &&&    0.90(1)         \\
Ref.~\cite{baek2013}(2013)               &&&&           -             &&&    0.9020(5)     \\
Ref.~\cite{kumano}(2013)                  &&&&     0.700(4)       &&&    0.904(5)       \\
Ref.~\cite{nishino_q6}(2016) 	        &&&&     0.70              &&&    0.88             \\
Ref.~\cite{JingChen}(2017)     	        &&&&     0.6658(5)     &&&    0.8804(2)     \\
Ref.~\cite{Chatterjee2018}(2018)      &&&&     0.681(1)       &&&    -                   \\
Ref.~\cite{Surungan2019}(2019)       &&&&     0.701(5)       &&&    0.898(5)       \\
Ref.~\cite{Hong2019}(2019)              &&&&     0.693            &&&    0.904           \\
Present work                                      &&&&     0.6901(4)     &&&    0.9127(5)     \\
\hline\hline
$q=7$                            		        &&&&    $T_{c1}$       &&&   $T_{c2}$       \\
\hline
Ref.~\cite{Borisenko2012}(2012)       &&&&     0.533           &&&    0.900           \\
Ref.~\cite{Chatterjee2018}(2018)       &&&&     0.531(6)      &&&    -                   \\
Present work                         	         &&&&     0.5305(3)     &&&    0.9071(5)    \\
\hline\hline
$q=8$                                 		 &&&&    $T_{c1}$       &&&   $T_{c2}$      \\
\hline	
Ref.~\cite{tomita}(2002)          	         &&&&     0.4259(4)     &&&    0.8936(7)    \\
Ref.~\cite{Baek2009}(2009)               &&&&     0.417(3)       &&&    0.894(1)      \\
Ref.~\cite{Chatterjee2018}(2018)       &&&&     0.418(1)       &&&    -                  \\
Present work                           		 &&&&     0.4172(3)     &&&    0.9060(5)   \\
\hline\hline
\end{tabular}
\label{Table:Tc_List}
\end{table}

\section*{C. $\mathbb{Z}_q$-deformed sine-Gordon theory}
The effective field theory describing the $q$-state clock model is the so-called $\mathbb{Z}_q$-deformed sine-Gordon model,
which is defined through its action~\cite{Wiegmann1978,Matsuo2006}
\begin{equation}
S=\frac{1}{2 \pi K} \int d^{2} \mathbf{r}(\nabla \phi)^{2}+\frac{g_{1}}{2 \pi \alpha^{2}} \int d^{2} \mathbf{r} \cos (\sqrt{2} \phi)
+ \frac{g_{2}}{2 \pi \alpha^{2}} \int d^{2} \mathbf{r} \cos (q \sqrt{2} \Theta),
\label{SEq:effectiveModel}
\end{equation}
where the real scalars $\phi$ and $\Theta$, being compactified on a circle as $\phi \equiv\phi+\sqrt{2} \pi$ (same for $\Theta$), are
mutually dual to each other, i.e., $\partial_{x} \phi=-K\partial_{y}\Theta$ and $\partial_{y} \phi=K\partial_{x}\Theta$.

The physics of the $q$-state clock model with $q=2,3,4$ can be easily understood from the above field theory~\cite{Lecheminant2002,LiWei2015}. For $q = 2$ and $3$, either of the
two cosine terms is relevant, and the model is thus generically off-critical. However, the self-dual point is very special and remains critical,
where the relevant perturbations drive the free-boson CFT (with central charge $c=1$) to new strong coupling fixed points (with $c < 1$
according to Zamolodchikov's $c$ theorem). For $q=2$, the cosine terms (having scaling dimension $1$) can be refermionized, and the
resulting theory at the self-dual point is a free massless Majorana fermion (Ising CFT with $c = 1/2$). For $q=3$, the self-dual case corresponds
to an integrable deformation of the $\mathbb{Z}_4$ parafermion CFT~\cite{Fateev1991,Lecheminant2002}, and the strong coupling fixed point
is the $\mathbb{Z}_3$ parafermion CFT (with $c = 4/5$). For $q = 4$, both cosine terms are marginally irrelevant at $K=4$ (otherwise one
of them is relevant). As there is only one critical point (manifestly self-dual via the Kramers-Wannier duality transformation) in the $q=4$
clock model, we can safely identify this with the field theory (\ref{SEq:effectiveModel}) with $K=4$ and $g_1 = g_2$, which is also self-dual
and, in fact, maps to two copies of Ising CFTs. In this way, the correspondence between the self-dual critical point of the $q$-state clock
model (\ref{SEq:qclock_Model}) for $q=2,3,4$ and the $\mathbb{Z}_q$-deformed sine-Gordon model (\ref{SEq:effectiveModel}) with
$K_{\mathrm{sd}}=q$ and $g_1=g_2$ is established.

For $q>4$, we have discussed the predictions of the above $\mathbb{Z}_q$-deformed sine-Gordon model in the main text.
Below we only supplement the derivation of the critical exponent $\eta=1/K$ for the spin-spin correlation function within the BKT critical phase. Within the critical phase, the effective action, after dropping the two irrelevant cosine terms,
corresponds to a free boson CFT
\begin{equation}
S=\frac{1}{2 \pi K} \int d^{2} \mathbf{r}(\nabla \phi)^{2}.
\label{SEq:Gaussian_model}
\end{equation}
For the free boson CFT~(\ref{SEq:Gaussian_model}), the two-point correlation function between $O_{m, n}= \exp (i m \sqrt{2} \phi) \exp (i n \sqrt{2} \Theta)$ and $O_{-m, -n}$ is given by~\cite{Matsuo2006}
\begin{equation}
\langle O_{m,n}(\vec{r_1}) O_{-m,-n}(\vec{r_2}) \rangle = \mathrm{exp} \left[-2\Delta_{m,n}\mathrm{log}\left(\frac{|\vec{r}_{12}|}{\alpha}\right)
- 2is_{m,n}\left(\mathrm{Arg}(\vec{r}_{12})+\frac{\pi}{2}\right)\right],
\label{SEq:Corr_func}
\end{equation}
where $\alpha$ is the ultraviolet cutoff, $\mathrm{Arg}(\vec{r}_{12})$ is the polar angle of the vector $\vec{r}_{12} = \vec{r_1} - \vec{r_2}$.
Here $\Delta_{m,n}$ and $s_{m,n}$ are the scaling dimension and the conformal spin of $O_{m, n}$,
\begin{equation}
\Delta_{m, n}(K)=\frac{1}{2}\left(m^{2} K+\frac{n^{2}}{K}\right), \quad s_{m,n} = mn,
\end{equation}
respectively.

To establish the correspondence between the above correlation function and the correlation function of the lattice spins, we should first know
the correspondence between the scalar fields and the lattice spin variables. From the symmetry analysis of the action~(\ref{SEq:effectiveModel}),
the scalar fields $\phi$ and $\Theta$ should be related to the vortex and lattice (clock) spins, respectively. Meanwhile, it is worth noting
that the field $\Theta$ has been compactified with $\Theta \equiv \Theta+\sqrt{2} \pi$, while the periodicity of the lattice spin is $2\pi$.
Thus, the clock spin on the lattice, in the continuum limit, can be identified with $O_{0,1}=\mathrm{exp}(i\sqrt{2} \Theta)$ (up to higher order terms with larger scaling dimensions).
According to Eq.~(\ref{SEq:Corr_func}), the critical exponent $\eta$ of the correlation function between the clock spins is therefore $\eta=2\Delta_{0,1}=1/K$.

\begin{figure*}[htb]
\centering
\includegraphics[width=17.5cm]{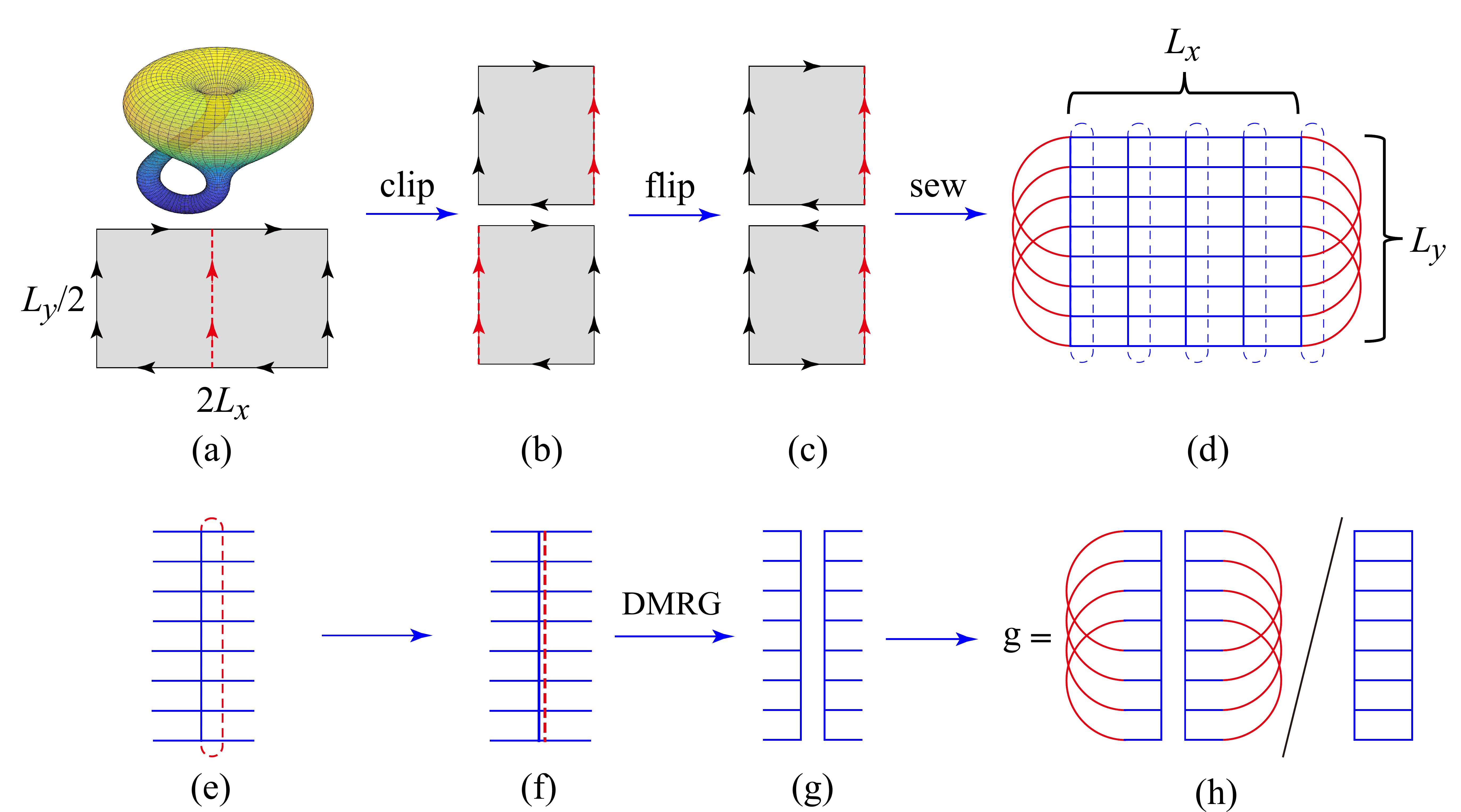}
\caption{The method for calculating the ratio between the Klein bottle and torus partition functions.
The Klein bottle partition function in (a) is converted into a cylinder with crosscap boundaries in (d) via (b) clip and (c) flip-and-sew
procedures. The bulk of the cylinder is defined via the column-to-column transfer matrix with (e) periodic and (f) open boundaries,
whose leading vector is approximated by an MPS by using the DMRG algorithm in (g).
(h) The Klein bottle ratio $g$ is obtained by contracting the MPS under properly defined boundaries.}
\label{Fig:Klein_approach}
\end{figure*}

\section*{D. Derivation of the boson radius $R=\sqrt{2K}$ for $q \ge 5$}

The recently proposed Klein bottle entropy approach~\cite{Tu2017,Tang2017a,Chen2017a,Wang2018,Tang2019a} provides a convenient way
for determining $K$ [see Eq.~(\ref{SEq:effectiveModel})] directly from the microscopic model (\ref{SEq:qclock_Model}).
It was shown in Ref.~\cite{Tang2019a} that the Klein bottle entropy is directly related to the boson radius $R$.
However, the calculation in Ref.~\cite{Tang2019a} was done with the convention in Ref.~\cite{yellowbook}.
Therefore, we have to translate the field theory (\ref{SEq:effectiveModel}) into the compactified boson
conformal field theory used in Ref.~\cite{yellowbook} [see Eq.~(6.40) with $g=1/4\pi$ and Eq.~(6.90)].

For this, making use of the dual relation
\begin{equation}
\partial_{x} \phi=-K\partial_{y}\Theta,\qquad \partial_{y} \phi=K\partial_{x}\Theta,
\end{equation}
we can rewrite (\ref{SEq:effectiveModel}) as
\begin{equation}
S =\frac{K}{2 \pi} \int d^{2} \mathbf{r}(\nabla \Theta)^{2}+\frac{g_{1}}{2 \pi \alpha^{2}} \int d^{2} \mathbf{r} \cos (\sqrt{2} \phi)
+\frac{g_{2}}{2 \pi \alpha^{2}} \int d^{2} \mathbf{r} \cos (q \sqrt{2} \Theta),
\label{Eq:DualTheory}
\end{equation}
where the fields $\phi$ and $\Theta$ are periodic
\begin{equation}
\phi \equiv \phi+\sqrt{2} \pi, \qquad
\Theta \equiv \Theta+\sqrt{2} \pi .
\end{equation}
Then, we rescale the fields $\phi$ and $\Theta$ in Eq.~(\ref{Eq:DualTheory}) as
\begin{equation}
\phi^{\prime} = \frac{1}{2 \sqrt{K}} \phi, \qquad \Theta^{\prime} = 2 \sqrt{K} \Theta,
\end{equation}
whose periodicity becomes
\begin{equation}
\phi^{\prime} \equiv \phi^{\prime}+ 2\pi / \sqrt{2K} , \qquad
\Theta^{\prime} \equiv \Theta^{\prime}+ 2\pi \sqrt{2K}.
\end{equation}
After the rescaling, the action in (\ref{Eq:DualTheory}) now takes the form
\begin{equation}
S =\frac{1}{8 \pi} \int d^{2} \mathbf{r}\left(\nabla \Theta^{\prime}\right)^{2}+\frac{g_{1}}{2 \pi \alpha^{2}} \int d^{2} \mathbf{r} \cos \left(2 \sqrt{2 K} \phi^{\prime}\right)
+ \frac{g_{2}}{2 \pi \alpha^{2}} \int d^{2} \mathbf{r} \cos \left(\frac{q}{\sqrt{2 K}} \Theta^{\prime}\right).
\end{equation}
Since both cosine terms are irrelevant for $T_{c1} < T < T_{c2}$, the effective theory is
\begin{equation}
S \simeq \frac{1}{8 \pi} \int d^{2} \mathbf{r}\left(\nabla \Theta^{\prime}\right)^{2},
\end{equation}
whose scalar field $\Theta^{\prime}$ has compactification radius $R = \sqrt{2K}$ [see also Eq.~(6.90) in Ref.~\cite{yellowbook}].

\section*{E. Klein bottle entropy approach}
Here we briefly review how to calculate the Klein bottle ratio~\cite{Tu2017,Tang2017a,Chen2017a,Wang2018,Tang2019a}
\begin{equation}
g = \frac{Z^{\mathcal{K}}\left(2 L_x, L_y/2\right)}{Z^{\mathcal{T}}(L_x, L_y)} ,
\label{SEq:Klein_Entropy}
\end{equation}
which directly gives the compactification radius $R=\sqrt{2K}$ for the compactified boson CFT~\cite{Tang2019a} or is
related to the quantum dimensions of the primary fields for the rational CFT~\cite{Tu2017}.

Fig.~\ref{Fig:Klein_approach}(a) corresponds to the Klein bottle partition function with length $2L_x$ and width $L_y/2$,
which is difficult to calculate directly. It is more convenient to convert it to a cylinder with crosscap boundaries [as
shown in Fig.~\ref{Fig:Klein_approach}(b-d)] by the clip, flip and sew procedures, where the spatial reflection symmetry of the
system has been used in the flip procedure. Since Eq.~(\ref{SEq:Klein_Entropy}) is valid for $L_x \gg L_y$,
the calculation of the Klein bottle partition function can be performed in the limit of $L_x \rightarrow \infty$,
for which only the leading eigenvector of the column-to-column transfer matrix is needed [see Fig.~\ref{Fig:Klein_approach}(e)].
In order to obtain this leading eigenvector, we first convert the periodic column-to-column
transfer matrix to a transfer matrix with open boundary [see Fig.~\ref{Fig:Klein_approach}(f)], and then use the DMRG algorithm
to obtain its leading eigenvector, which is approximated by a MPS with bond dimension $D$. Once the MPS
is obtained, the ratio of the Klein bottle and torus partition functions can be calculated as shown in Fig.~\ref{Fig:Klein_approach}(h).

Fig.~\ref{Fig:Klein_q234} shows the Klein bottle ratio $g$ for the clock models with $q=2,3,4$. It is clearly seen that the curves for different $L_y$ intersect at the self-dual points, where second-order phase transitions occur. For these second-order transition points, the
Klein bottle ratios $g$ at the critical points of the $q=2,3$ and $4$ clock model are related to the quantum dimensions of the primary fields~\cite{Tu2017,Tang2017a}.

\begin{figure*}[htp]
\centering
\includegraphics[width=17.5cm]{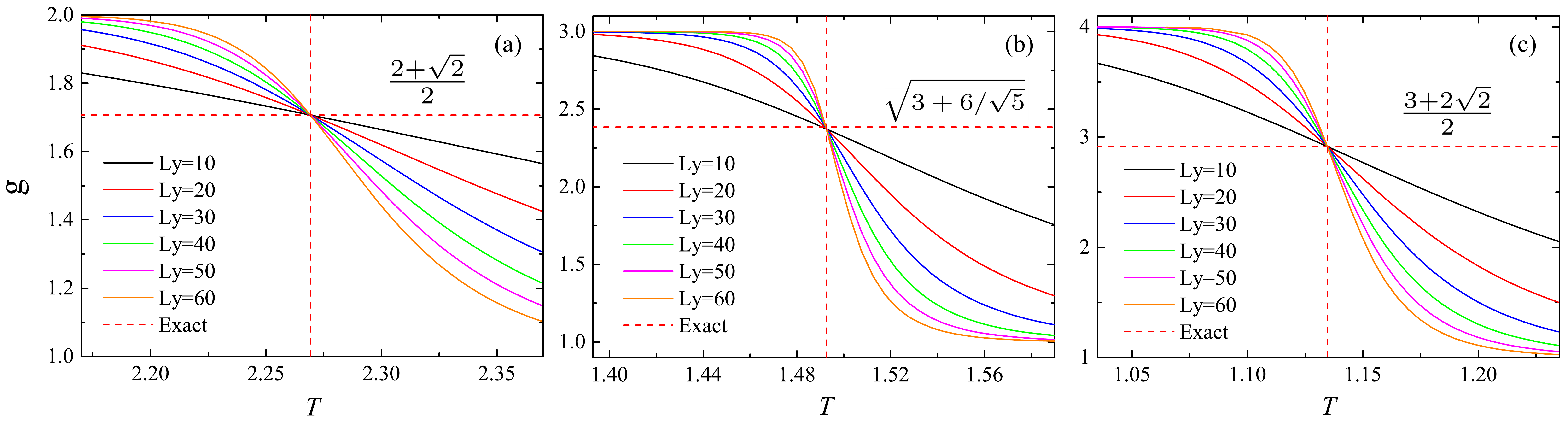}
\caption{The Klein bottle ratios $g$ as a function of temperature $T$ for (a) the Ising model ($q=2$), (b) $q=3$ clock model, and (c) $q=4$ clock model.
The different curves correspond to calculations with different $L_y$. The vertical dashed lines indicate the known second-order transition points. The horizontal lines are the Klein bottle ratios at the critical points, which are predicted by CFTs~\cite{Tu2017,Tang2017a}.
}
\label{Fig:Klein_q234}
\end{figure*}

\end{widetext}

\end{document}

%% file: preamble.tex
\usepackage{braket}
\usepackage{graphicx}
\usepackage[english]{babel}
\usepackage{color}
\usepackage{tabularx}
\usepackage{algorithm}
\usepackage{algpseudocode}
\usepackage{bm}
\usepackage{tikz}
\usepackage{subfigure}
\usepackage[pdfpagelabels,plainpages=false,bookmarks=true,colorlinks,linkcolor=red,urlcolor=blue,citecolor=blue]{hyperref}   

\newcommand{\mpsA}[3]{\draw (#1,#2) circle (.5); \draw (#1,#2) node {$#3$};}
\newcommand{\mpsAL}[3]{\draw (#1+0.5,#2) --(#1+0.25,#2+0.5) --(#1-0.5,#2+0.5) --(#1-0.5,#2-0.5) --(#1+0.25,#2-0.5) --cycle; \draw (#1,#2) node {$#3$};}
\newcommand{\mpsAR}[3]{\draw (#1-0.5,#2) --(#1-0.25,#2+0.5) --(#1+0.5,#2+0.5) --(#1+0.5,#2-0.5) --(#1-0.25,#2-0.5) --cycle; \draw (#1,#2) node {$#3$};}

\newcommand{\mpsAC}[3]{\draw[rounded corners] (0.5+#1,0.5+#2) rectangle (-0.5+#1,-0.5+#2); \draw (#1,#2) node {$#3$};}
\newcommand{\mpoT}[3]{\draw (0.5+#1,0.5+#2) rectangle (-0.5+#1,-0.5+#2); \draw (#1,#2) node {$#3$};}
\newcommand{\centerC}[3]{\draw[rounded corners] (0.5+#1,0.5+#2) rectangle (-0.5+#1,-0.5+#2); \draw (#1,#2) node {$#3$};}

\newcommand{\lineH}[3]{\draw (#1,#3) -- (#2,#3);}
\newcommand{\lineV}[3]{\draw (#3,#1) -- (#3,#2);}

\newcommand{\lineCul}[2]{\draw[rounded corners] (#1,#2-0.5) --(#1,#2) --(#1+0.5,#2);}
\newcommand{\lineCdl}[2]{\draw[rounded corners] (#1,#2+0.5) --(#1,#2) --(#1+0.5,#2);}
\newcommand{\lineCur}[2]{\draw[rounded corners] (#1,#2-0.5) --(#1,#2) --(#1-0.5,#2);}
\newcommand{\lineCdr}[2]{\draw[rounded corners] (#1,#2+0.5) --(#1,#2) --(#1-0.5,#2);}

\newcommand{\cornerIultens}[2]{
\lineCul{#1}{#2};
\lineH{#1+0.5}{#1+1.0}{#2};
\lineV{#2-0.5}{#2-1.0}{#1};
}

\newcommand{\cornerIdltens}[2]{
\lineCdl{#1}{#2};
\lineH{#1+0.5}{#1+1.0}{#2};
\lineV{#2+0.5}{#2+1.0}{#1};
}

\newcommand{\cornerIurtens}[2]{
\lineCur{#1}{#2};
\lineH{#1-0.5}{#1-1.0}{#2};
\lineV{#2-0.5}{#2-1.0}{#1};
}

\newcommand{\cornerIdrtens}[2]{
\lineCdr{#1}{#2};
\lineH{#1-0.5}{#1-1.0}{#2};
\lineV{#2+0.5}{#2+1.0}{#1};
}

\newcommand{\centerIudtens}[2]{
\lineH{#1-0.5}{#1+0.5}{#2};
\lineH{#1-0.5}{#1-1.0}{#2};
\lineH{#1+0.5}{#1+1.0}{#2};
}

\newcommand{\MPOtens}[3]{
\mpoT{#1}{#2}{#3};
\lineH{#1-0.5}{#1-1.0}{#2};
\lineH{#1+0.5}{#1+1.0}{#2};
\lineV{#2+0.5}{#2+1.0}{#1};
\lineV{#2-0.5}{#2-1.0}{#1};
}

\newcommand{\MPSuLtens}[3]{
\mpsAL{#1}{#2}{#3};
\lineH{#1-1.0}{#1-0.5}{#2};
\lineH{#1+0.5}{#1+1.0}{#2};
\lineV{#2-1.0}{#2-0.5}{#1};
}

\newcommand{\MPSuRtens}[3]{
\mpsAR{#1}{#2}{#3};
\lineH{#1-0.5}{#1-1.0}{#2};
\lineH{#1+0.5}{#1+1.0}{#2};
\lineV{#2-0.5}{#2-1.0}{#1};
}

\newcommand{\MPSuCtens}[3]{
\mpsAC{#1}{#2}{#3};
\lineH{#1-0.5}{#1-1.0}{#2};
\lineH{#1+0.5}{#1+1.0}{#2};
\lineV{#2-0.5}{#2-1.0}{#1};
}

\newcommand{\MPSrtens}[3]{
\mpsA{#1}{#2}{#3};
\lineH{#1-0.5}{#1-1.0}{#2};
\lineV{#2-0.5}{#2-1.0}{#1};
\lineV{#2+0.5}{#2+1.0}{#1};
}

\newcommand{\MPSdLtens}[3]{
\mpsAL{#1}{#2}{#3};
\lineH{#1-0.5}{#1-1.0}{#2};
\lineH{#1+0.5}{#1+1.0}{#2};
\lineV{#2+0.5}{#2+1.0}{#1};
}

\newcommand{\MPSdRtens}[3]{
\mpsAR{#1}{#2}{#3};
\lineH{#1-0.5}{#1-1.0}{#2};
\lineH{#1+0.5}{#1+1.0}{#2};
\lineV{#2+0.5}{#2+1.0}{#1};
}

\newcommand{\MPSdCtens}[3]{
\mpsAC{#1}{#2}{#3};
\lineH{#1-0.5}{#1-1.0}{#2};
\lineH{#1+0.5}{#1+1.0}{#2};
\lineV{#2+0.5}{#2+1.0}{#1};
}

\newcommand{\MPSltens}[3]{
\mpsA{#1}{#2}{#3};
\lineH{#1+0.5}{#1+1.0}{#2};
\lineV{#2-0.5}{#2-1.0}{#1};
\lineV{#2+0.5}{#2+1.0}{#1};
}

\newcommand{\centerCudtens}[3]{
\centerC{#1}{#2}{#3};
\lineH{#1-0.5}{#1-1.0}{#2};
\lineH{#1+0.5}{#1+1.0}{#2};
}

